\documentclass[11pt,onecolumn,letterpaper,final]{IEEEtran}
\usepackage{epsfig,graphicx}
\usepackage{amssymb,amsmath}
\usepackage[caption = false, singlelinecheck = false]{subfig}
\usepackage{url}

\newcommand {\ul} {\underline}

\newcommand {\trm} {\textrm}
\newcommand {\BMAT}{\begin{matrix}}
\newcommand {\EMAT}{\end{matrix}}
\begin{document}

\title{Cooperative localization using angle of arrival measurements: sequential algorithms and non-line-of-sight suppression}

\author{Bharath~Ananthasubramaniam and~Upamanyu~Madhow\\
        Electrical and Computer Engineering Department\\
        University of California, Santa Barbara\\
        Santa Barbara, CA 93106\\
        E-mail: bharath@engr.ucsb.edu, madhow@ece.ucsb.edu} \maketitle
\vspace{-10mm}
\begin{abstract}
We investigate localization of a source based on angle of arrival (AoA) measurements made at a geographically dispersed network of cooperating receivers. The goal is to efficiently compute accurate estimates despite \emph{outliers} in the AoA measurements due to multipath reflections in non-line-of-sight (NLOS) environments. Maximal likelihood (ML) location estimation in such a setting requires exhaustive testing of estimates from all possible subsets of ``good'' measurements, which has exponential complexity in the number of measurements.  We provide a randomized algorithm that approaches ML performance with linear complexity in the number of measurements.  The building block for this algorithm is a low-complexity sequential algorithm for updating the source location estimates under line-of-sight (LOS) environments.  Our Bayesian framework can exploit the ability to resolve multiple paths in wideband systems to provide significant performance gains over narrowband systems in NLOS environments, and easily extends to accommodate additional information such as range measurements and prior information about location.
\end{abstract}

\section{Introduction}
In this paper, we investigate cooperative localization of a source transmitting a known signal using a network of geographically dispersed receivers (detectors) using angle of arrival (AoA) measurements. We consider AoA measurements, since they only require that each receiver has a calibrated antenna array with known orientation and location, and that the receivers can coordinate to pool all the AoA estimates corresponding to a given source (e.g., using coarse timing synchronization between the receivers to associate AoAs for the signal from a given source at a given point of time). While the geometry of AoA-based source localization is straightforward for line-of-sight (LOS) environments, our primary goal in this paper is to develop efficient localization algorithms for non-line-of-sight (NLOS) environments.

While the problem considered here is of broad applicability, our primary motivation is localization in sensor networks, where a network of collector nodes (i.e., data gathering nodes with some advanced capabilities such as AoA estimation) collaboratively decode and localize transmissions from elementary microsensors that have data to send. Since sensors communicate only when they observe an ``interesting'' event, without prior coordination with the collector nodes, this \emph{sensor-driven} paradigm (proposed in \cite{Ananthasubramaniam2007}), allows drastic reduction in the microsensor functionality and communication energy costs. By locating the sensors transmitting in response to the event, the location of the event can be estimated.  Our goal, therefore, is to locate the transmitting sensor in NLOS propagation conditions. Note that the term `sensor' is not a reference to receivers, as is sometimes the usage in the source localization and array processing literature.  To avoid confusion, we use the term ``source'' to refer to the source of the transmitted signal, and the term ``receiver'' for a device that receives this signal and produces an AoA estimate.  This problem is also applicable to asset tracking using active radio frequency identification (RFID) tags, in which tags periodically or intermittently transmit a signal to be used to localize them \cite{aeroscout,savi,wherenet}.

In the preceding scenario, other measurements such as time-difference-of-arrival (TDOA) \cite{aeroscout} and received signal strength (RSS) \cite{ekahau} could also be used for localization. TDOA measurements require stringent synchronization between the receivers, while RSS measurements are highly sensitive to the propagation environment, and typically require extensive calibration. On the contrary, AoA measurements offer the prospect of providing accurate localization without tight coordination among receivers or extensive calibration, at least in LOS environments.  In this paper, we investigate whether this promise can be realized in more realistic NLOS environments, with the understanding that AoA-based performance can be augmented by incorporating information from TDOA and RSS measurements, if available.

We previously proposed one possible receiver design (including timing acquisition and array processing algorithms) to generate these AoA estimates \cite{Ananthasubramaniam2007}.  In this paper, we abstract away such details in order to focus on the fundamentals of AoA-based localization.  To this end, we propose statistical models for AoA measurements that model the performance of a broad class of AoA estimators in the literature under a variety of propagation environments.  These models permit performance comparisons between our proposed algorithms and fundamental limits such as the Cramer-Rao Lower Bound (CRLB). Our main contributions are as follows:

\begin{enumerate}

\item 
As a building block for localization in NLOS environments, we develop a \emph{scalable sequential algorithm} for aggregation of AoA estimates from multiple receivers to estimate the source location in LOS scenarios (minimal multipath scattering). The receivers only need to exchange source location and covariance estimates, where the source location estimate is updated with the local AoA measurement before it is forwarded to the next receiver. This approximately maximum likelihood (ML) algorithm has linear complexity in the number of receivers. We use the CRLB to estimate the location uncertainty as a function of the coverage area, the variance of the AoA measurements, and the number of receivers.

\item 
We model the AoA estimates due to multipath components with AoA measurements far away from the LOS path as {\it outliers.}  Assuming that at least some of the receivers see near-LOS paths, ML localization requires brute force elimination of outliers by considering all possible subsets of AoA measurements, which is excessively complex. We propose a randomized algorithm with $O(MN^2)$ complexity for \emph{outlier suppression}, which employs $M$ randomly initialized instances of the sequential algorithm.  Here $M$ is chosen to achieve a given probability of localization failure, and does not grow with $N$, the number of receivers.  

\item 
The proposed algorithms are numerically shown to achieve the CRLB and the ML performance. We quantify the performance penalty due to the presence of outliers, and provide examples that illustrate the performance advantage of wideband systems, which can resolve LOS and NLOS paths, relative to narrowband systems without the capability for providing such resolution. Also, our Bayesian framework permits integration of other sensing modalities such as RSS-based range measurements for localization in a NLOS environment.
\end{enumerate}
An outline of these preceding results was presented previously at a workshop \cite{melt08}. This paper significantly expands upon that work, including detailed derivations of the algorithms and performance bounds, as well as a more comprehensive set of simulation results.

{\bf Related Work:} There is a rich history of literature on source localization using a variety of measurements, including TDOA, AoA and RSS that depend on LOS channels between the source and the receivers with no multipath and consider errors due to measurement noise alone. A detailed discussion of the body of standard localization techniques using multiple modalities, their limitations and practical considerations are presented in \cite{Patwari2005,Mao2007}. 

Our approach to localization in NLOS environments draws upon the significant literature on handling outliers \cite{robust}. One approach is to identify and remove outliers {\it before} location estimation. In \cite{borras}, a generalized likelihood ratio test is used to identify NLOS estimates; this assumes at least partial statistical knowledge about the NLOS components and does not utilize the inherent consistency between measurements corresponding to a single source. Prior knowledge of the NLOS characteristics is used to eliminate outliers in TDOA and AoA measurements in \cite{Cong2005}, but requires exploring all subsets of measurements, which is not scalable. The skewed distribution of range or time of flight (TOF) estimates due to a positive bias introduced by NLOS propagation is used in \cite{venkat} to identify and correct for NLOS errors. Correction of NLOS time of arrival estimates prior to localization is accomplished by Kalman filtering in \cite{bao} and soft combining of pseudo-range measurements in \cite{seshan}. The residuals from locating the source using a least-squares algorithm on a set of AoA measurements are compared against a threshold to identify and eliminate outliers in \cite{xiong}. More recently, Yu et al. \cite{Yu2009} formulate a Neyman-Pearson test to only identify the outliers in different combinations of localization measurements such as AoA, ToA and RSS-based ranging and similarly, G\"{u}ven\c{c} et al. \cite{Guevenc2008} identify NLOS components in multipath ultrawideband (UWB) channel metrics using joint likelihood ratio tests. Al-Jazzar et al. \cite{Al-Jazzar2009} design a nonlinear optimization algorithm with nonlinear constraints to compute the locations of the unknown scatterers to identify the NLOS components. G\"{u}ven\c{c} et al. present a detailed survey of UWB TOA-based NLOS mitigation techniques in \cite{Guvenc2009}.

In this paper, we adopt the alternative approach of robust location estimation while simultaneously limiting the effect of outliers (called NLOS ``mitigation"), thus eliminating outliers ``on the go.'' To the best of our knowledge this approach has only once been used with AoA measurements in \cite{Tang2008} and similar work for range/TOF measurements includes \cite{chen,Venkatesh2007,casas}. Tang et al. \cite{Tang2008} formulate the localization using AoA and ToA as an optimization with a geometrically-constrained objective function that has a higher computational complexity of at least $O(N^3)$. More significantly, they only consider angular spreads due to a disk of scatterers around the transmitter that results in a simple Gaussian model of AoA spread. Thus, this AoA distribution is equivalent to having Gaussian errors in the LOS estimate and does not account for blockage of LOS. Chen \cite{chen} proposes a weighted least-squares localization using range/TOF measurements, where the weights are iteratively varied to assign the lowest weights to outliers and highest to the LOS estimates. The main drawback of this approach is the $O(2^N)$ complexity due to a combinatorial exploration of subsets of ``good'' measurements, which does not scale. Venkatesh and Buehrer \cite{Venkatesh2007} present a linear programming approach (of approximately $O(N^3)$ complexity) to mitigate NLOS using ToA estimates in a UWB system that utilizes the positive bias in NLOS range measurements.

Our approach shares many similarities with Casas et al. \cite{casas}. Casas et al. perform closed-form trilateration of random triplets of measurements (similar to the $M$ randomizations in our method) followed by a selection of the location estimate with the smallest median residual, which has an $O(MN)$ complexity. They propose a slight improvement by performing a multilateration after identification of the outliers from the previous step, incurring an $O(N^2)$ complexity. However, localization even without outliers using AoA measurements does not have a closed form solution in terms of more than two noisy AoA measurements in two-dimensions (and more than three noisy AoA measurements in three-dimensions), i.e., in an over specified system, due to the trigonometric dependence of AoA on the source location. This nonlinear dependence of AoA on the source location makes the algorithmic complexity inherently larger than for the time of flight measurements considered in \cite{casas}.  Our sequential algorithm offers a low-complexity approach to localization, which can be viewed as a generalization of \cite{casas}: we naturally grow ``good'' subsets of measurements to include many LOS measurements, without needing a separate phase for localization using inliers. Furthermore, our Bayesian framework permits incorporation of other localization modalities, such as RSS-based range measurements, with little change to the algorithm (see Figure \ref{fig:LOSrangeinfo}).  
  
The rest of the paper is organized as follows. AoA measurement models under different propagation environments are described in Section \ref{sec:models}. The sequential algorithm for localization in LOS scenarios, along with performance benchmarks, are presented in Section \ref{sec:LOSscen}. We derive an extension to the sequential localization to suppress NLOS AoA estimates in Section \ref{sec:NLOSalg}, which is motivated by the structure of the ML estimate. The proposed algorithms are numerically investigated in Section \ref{sec:sims}.  Section \ref{sec:concl} contains concluding remarks.

\section{Models of AoA Estimates}\label{sec:models}

Each receiver estimates the direction of arrival of the source transmission using an antenna array.  In this section, we discuss models for the estimated AoA that capture the effects of LOS and NLOS propagation environments. These models are then used to design source localization algorithms in Sections \ref{sec:LOSscen} and \ref{sec:NLOSalg}.  For simplicity, we assume a linear array, although our framework generalizes to arbitrary antenna arrays, as long as the array manifold is known.

Classical AoA estimation techniques such as MUSIC and ESPRIT \cite{godara} are designed to separate uncorrelated point sources assuming LOS propagation from the source to the receivers. This point-source propagation model does not account for multipath scattering encountered in many deployment environments. Multipath adds highly correlated AoA components to the received signal, and extensions to deal with such effects using techniques such as spatial smoothing have been proposed at the cost of lower AoA resolutions and poorer source separation capabilities. 
 
Here, we adopt the approach taken in \cite{raich,ertel,meng,trump,valaee}, where a series of models for AoA estimation in the presence of scattering have been proposed.  The propagation is characterized by a mean arrival angle (corresponding to the true bearing) and a spatial spreading parameter (quantifying the spatial uncertainty caused by multipath). The resulting AoA at the antenna array is a random variable, typically modeled as Gaussian \cite{trump} or Laplacian \cite{spencer}.  Drawing on these ideas, we propose the following models:

 {\bf LOS model:} We characterize the spatial spreading in LOS propagation scenarios by zero-mean symmetric finite variance ``noise" models, such as the Gaussian and Laplacian. AoA estimation under LOS in the presence of additive white Gaussian noise results in zero-mean Gaussian errors, whose variance depends on the signal-to-noise ratio \cite{rao_hari}. Additional errors in the AoA measurements due to local scattering in the vicinity of the source are also modeled using zero-mean symmetric distributions. Therefore, Gaussian and Laplacian error models seamlessly transition between scenarios with and without local scattering for different values of the spatial spreading parameter. These models represent situations where the received signal has a strong LOS component, together with a limited amount of scattering. For a source at location $\ul{X}$ along a true bearing $\theta(\ul{X})$, the Gaussian LOS model with local scattering is
  \begin{equation}\label{eqn:LOSmodel}
   p_\trm{Gaus}(\hat{\theta}/\ul{X}) = \frac{1}{\sqrt{2\pi}\sigma(1-2Q(\frac{\pi}{2\sigma}))}\exp\left(-\frac{(\hat{\theta} - \theta(\ul{X}))^2}{2\sigma^2}\right),~~\hat{\theta} \in \left[-\frac{\pi}{2},\frac{\pi}{2}\right],
  \end{equation}
 where $\sigma^2$ represents the spatial extent of scattering, $Q(t) = \int_{t}^{\infty}\exp(-t^2/2)dt/\sqrt{2\pi}$ is the normal tail distribution and all angles are measured with respect to the antenna broadside. The permissible angles are within $\pm \pi/2$ of the antenna broadside and the Gaussian density is truncated to reflect this.
 As the spatial spread $\sigma^2$ increases, the density become progressively long tailed and tends toward a uniform density. The Laplacian LOS model with spatial spreading factor $\sigma$ has heavier tails than the Gaussian model:
 \begin{equation}\label{eqn:Lapmodel}
   p_\trm{Lap}(\hat{\theta}/\ul{X}) = \frac{1}{2\sqrt{2}\sigma \exp(-|\pi/(\sqrt{2}\sigma)|)}\exp\left(-\frac{\sqrt{2}|\hat{\theta} - \theta(\ul{X})|}{\sigma}\right),~~\hat{\theta} \in \left[-\frac{\pi}{2},\frac{\pi}{2}\right].
 \end{equation}

{\bf LOS blockage model:} Environments where the LOS path to a receiver is blocked by structures, such as building, hills or trees, are also relevant. As a result, the received signal is composed exclusively of multipath components that deviate ``far" from the LOS path. Therefore, AoA measurements made from these scattered and reflected paths alone are fairly uncorrelated with the true bearing of the source. An AoA estimate drawn uniformly from the feasible set aptly models such a scenario:
 \begin{equation}\label{eqn:blockmod}
    p_\trm{NLOS} (\hat{\theta}/\ul{X}) = \frac{1}{\pi}, ~~\hat{\theta} \in \left[-\frac{\pi}{2},\frac{\pi}{2}\right].
 \end{equation}

When there are significant contributions from both the LOS path and the multipath components from other directions, the model for the AoA estimates depends on the capability of the receiver to resolve these contributions spatially. If the source signal has a large enough bandwidth, the multipath components can be resolved in time, and separate AoA estimates can be obtained for each path.  If the bandwidth is insufficient to resolve the paths, then the receiver may only be able to obtain a single AoA estimate. We model these two scenarios separately.

 {\bf Narrowband multipath model:}
 With narrowband source transmissions, the receiver is capable of resolving the arriving combination of LOS path and NLOS multipath in the spatial domain only. For receivers with relatively small number of antenna elements, the receiver can only measure the AoA of the strongest arriving path (or strongest superposition of paths). Hence, each receiver produces a single AoA estimate that, depending on the relative strengths of the LOS and NLOS components, will provide an estimate close to or very ``far" from the true bearing of the source (outliers). Accordingly, we model a typical narrowband scenario as follows: Let $\alpha$ be the fraction of receivers that experience reflected and scattered multipath significantly stronger than the LOS path. In this event, the AoA appears to correspond to the LOS path being blocked and hence, is drawn from the worst-case NLOS model in \eqref{eqn:blockmod}. In the remaining instances, when the LOS path is strong, the AoA is drawn from one of the LOS models, for instance \eqref{eqn:LOSmodel}. Thus, we arrive at the following narrowband model:
\begin{equation} 
   p_\trm{narrowband}(\hat{\theta}/\ul{X}) = \alpha~p_\trm{NLOS}(\hat{\theta}/\ul{X}) + (1-\alpha)~p_\trm{Gaus}(\hat{\theta}/\ul{X}).\label{eqn:narrowbndmod}
\end{equation}
 This model also represents the least favorable multipath environment, where the LOS component is either present and significant, or is blocked (completely absent), and serves as our candidate model for developing the NLOS suppression algorithm. Note that distributions with heavy-tails, such as the Laplacian (in \eqref{eqn:Lapmodel}) or a Cauchy model, can also be used to represent the narrowband setting with outliers. We investigate these alternate models in Section \ref{sec:NLOSperf}.

 {\bf Wideband multipath model:}
With wideband source transmissions, the receiver has an additionally resolve paths in time as well. When the scattered or reflected paths are sufficiently temporally or spatially separated from the LOS path, the receiver is capable of resolving the LOS path, and possibly multiple NLOS paths.  We model the multiple AoA estimates produced by the receiver as follows: the AoA estimate corresponding to the LOS path is represented by an LOS model such as \eqref{eqn:LOSmodel} or \eqref{eqn:Lapmodel}, while the remaining estimates are drawn uniformly from the feasible set, as in the worst-case model with LOS blockage. Thus, if a receiver resolves $L$ paths, the resulting AoA estimates are generated according to the following distribution:
 \begin{equation}
   p_\trm{wideband}(\hat{\theta^{(1)}},\hat{\theta^{(2)}}\ldots \hat{\theta^{(L)}}/\ul{X}) =~ p_\trm{Gaus}(\hat{\theta^{(1)}}/\ul{X})~ p_\trm{NLOS}(\hat{\theta^{(2)}}/\ul{X})\ldots ~p_\trm{NLOS}(\hat{\theta^{(L)}}/\ul{X}).
 \end{equation}
Assuming that the LOS path is not blocked, it is resolved by a receiver in the wideband
scenario.  However, it is not known {\it a priori} which of the $L$ AoA estimates corresponds to the LOS path. If the LOS path is blocked, then none of the $L$ resolved AoA estimates might be close to the true source bearing.
\begin{figure*}[ht!]
  \includegraphics[width=6.5in,clip]{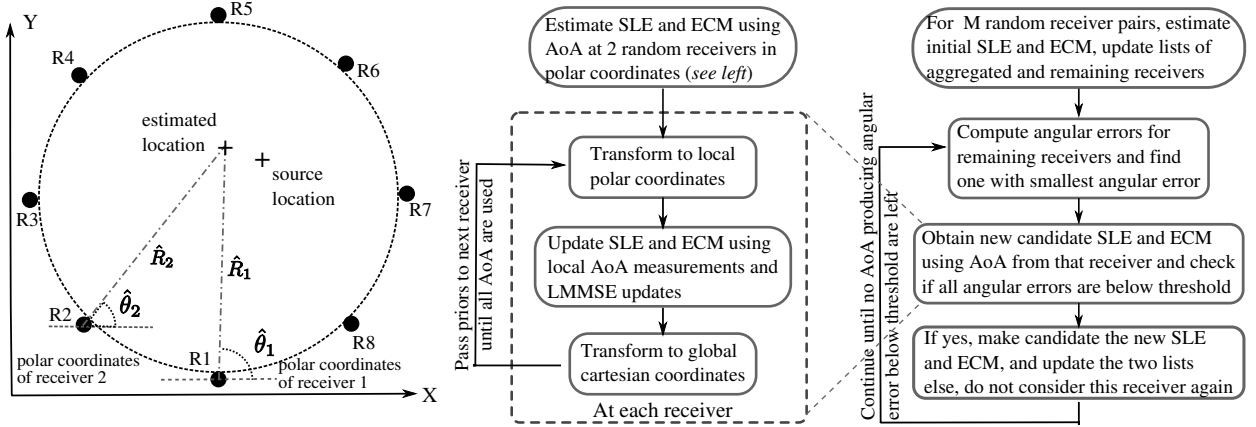}
  \centering
  \caption{(\emph{Left panel}) One possible sensor network layout with a circular field and 8 receivers on its perimeter: The geometry used to compute the ``bootstrap'' source estimate from AoA estimates of receivers R1 and R2 is shown. (\emph{Center Panel}) The flowchart of the sequential localization algorithm using AoA measurements (with no outliers) presented in Section \ref{sec:localg}. (\emph{Right Panel}) The flowchart of the outlier suppresion algorithm in Section \ref{sec:aggr_outlierdet} that uses the sequential algorithm in the \emph{Center Panel.}}\label{fig:bootstrap}
 \end{figure*}
 
\section{Localization in LOS scenarios}\label{sec:LOSscen}

In this section, we consider scenarios with only LOS propagation, i.e., no LOS blockage, where the spread in the AoA estimates is caused only by local scattering. We present an algorithm for sequential aggregation of the available AoA estimates (generated according to the LOS model in Section \ref{sec:models}) to produce the estimate of the source location. Each receiver performs linear minimum mean squared error (LMMSE) updates on the \emph{prior} source location estimate (SLE) (received from a previous receiver) using its own AoA estimate, before passing the updated estimate to the next receiver. This process is continued until all the available AoA estimates is aggregated into the SLE. 

\subsection{LMMSE Updates and Coordinate Transformations}\label{sec:comb_tform}

 At each receiver, we desire linear updates (to keep the computational burden to a minimum) for arbitrary measurement models, while also propagating only the first- and second-order statistics of the sensor location. Both these requirements are satisfied by an LMMSE estimator. Working with AoA measurements at each receiver, the SLE updates are conveniently formulated in the polar coordinates $[R~\theta]$ centered at that receiver, where $R$ is the distance between the source and receiver, and $\theta$ is measured from the x-axis (see Figure \ref{fig:bootstrap}, Left Panel). Additionally, this choice of polar coordinates makes the LMMSE update optimal under Gaussian measurement models such as the LOS model in Section \ref{sec:models}.

 Each receiver receives the following prior information: SLE $\underline{\bar{\mu}} = [\bar{R}~~\bar{\theta}]^T$ and ECM $\bar{\mathbf\Sigma}$, where
  $$
  \mathbf{\bar{\Sigma}} = \left(\begin{array}{cc}
                                     \bar{\Sigma}_{RR} &  \bar{\Sigma}_{R\theta}\\
                                     \bar{\Sigma}_{R\theta} & \bar{\Sigma}_{\theta\theta}
                                     \end{array}\right).
 $$
 The AoA estimate at the receiver is $\tilde{\theta}$ with spatial spread $\tilde{\sigma}_{\theta}$. Note, henceforth, we use the convention that new measurements are represented by $\tilde{.}$, prior information by $\bar{.}$, update variables by $\hat{.}$, vectors by $\underline{.}$ and matrices by bold typeface. We desire a linear SLE update,
 \begin{equation}\label{eqn:linupdate}
    \underline{\hat{\mu}} = \underline{\bar{\mu}} + \underline{K}(\tilde{\theta} - A\underline{\bar{\mu}}),
 \end{equation}
 where $\underline{K}$ is the Kalman gain and $A = [0~1]$ (only new AoA estimates are available at the receiver). For the innovation $(\tilde{\theta} - A\bar{\mu})$ to be orthogonal to the estimate $\hat{\mu}$, the Kalman gain must be:
 \begin{equation}\label{eqn:kalgain}
     \underline{K} = A\mathbf{\bar{\Sigma}}~(\tilde{\sigma}_{\theta} + A\mathbf{\bar{\Sigma}}A^T)^{-1}.
 \end{equation}
Therefore, inserting \eqref{eqn:kalgain} into \eqref{eqn:linupdate}, the updated SLE, $\hat{\mu}$, is obtained as
 \begin{eqnarray}
 \hat{R} &=& \bar{R} +
 \frac{\bar{\Sigma}_{R\theta}(\tilde{\theta}-\bar{\theta})}{\bar{\Sigma}_{\theta\theta}+\hat{\sigma}_{\theta}},\nonumber\\
 \hat{\theta} &=& \frac{\bar{\Sigma}_{\theta\theta}~\tilde{\theta}+\tilde{\sigma}_{\theta}~\bar{\theta}}{\bar{\Sigma}_{\theta\theta}+\tilde{\sigma}_{\theta}}.\label{eqn:Meanupdate}
 \end{eqnarray}
 The ECM update can be obtained using \eqref{eqn:linupdate} as
 \begin{equation}\label{eqn:cov_mat_update}
   \mathbf{\hat{\Sigma}}^{-1} = \mathbf{\bar{\Sigma}}^{-1} + \left(
                                                                                     \begin{array}{cc}
                                                                                       0 & 0 \\
                                                                                       0 & ~(\tilde{\sigma}_{\theta})^{-1}\\
                                                                                     \end{array}
                                                                                   \right).
 \end{equation}
The entries of the updated ECM are given by
 \begin{eqnarray}
 \hat{\Sigma}_{RR} &=& \bar{\Sigma}_{RR} - \frac{(\bar{\Sigma}_{R\theta})^2}{\bar{\Sigma}_{\theta\theta}+\tilde{\sigma}_{\theta}},\nonumber\\
 \frac{1}{\hat{\Sigma}_{\theta\theta}} = \frac{1}{\bar{\Sigma}_{\theta\theta}}  + \frac{1}{\tilde{\sigma}_{\theta}}~~&\textnormal{and}&~~\hat{\Sigma}_{R\theta}=\frac{\tilde{\sigma}_{\theta}~\bar{\Sigma}_{R\theta}}{\bar{\Sigma}_{\theta\theta}+\tilde{\sigma}_{\theta}}.\label{eqn:Covupdate}
 \end{eqnarray}
 When the observations $\tilde{\theta}$ have Gaussian errors as in the LOS model \eqref{eqn:LOSmodel}, the LMMSE updates in \eqref{eqn:Meanupdate} are also the optimal minimum mean squared error updates. Under non-Gaussian error models, the LMMSE updates are the optimal linear updates from the perspective of mean squared error.

The updated SLE $\underline{\hat{\mu}}$ and ECM $\mathbf{\hat{\Sigma}}$ are thus produced in a polar coordinate system centered at the receiver with the new measurement. This SLE and ECM must then be transformed to the global cartesian system (a common frame of reference) to provide the information in a form accessible to the next receiver.  The next receiver then transforms these estimates into its own polar coordinate system.  We now describe transformations between the local polar coordinate and global cartesian coordinate systems.

 Suppose that the receiver located at $\underline{X}_c = [x_c~y_c]^T$ in the global cartesian system computes the new SLE $\underline{\hat{\mu}} = [\hat{R}~~\hat{\theta}]^T$ with ECM $\hat{\mathbf \Sigma}_\trm{pol}$ in polar coordinates. The SLE in the cartesian system, $\underline{X}_s = [x_s~y_s]^T$, is given by 
 \begin{equation}
\underline{X}_s = \underline{X}_c + [\hat{R}\cos(\hat{\theta})~~\hat{R}\sin(\hat{\theta})].\label{eqn:CartEst}
 \end{equation}
 Therefore, errors in $[\hat{R}~~\hat{\theta}]^T$ can be mapped to errors in $[x_s~y_s]^T$ as
 $$
   \left(\begin{array}{c}
    dx_s\\dy_s\end{array}\right) = \overbrace{\left(\begin{array}{cc} \cos(\hat{\theta}) &
-\hat{R}\sin(\hat{\theta})\\\sin(\hat{\theta}) & \hat{R}\cos(\hat{\theta})
\end{array}\right)}^{\displaystyle T_\trm{pol}(\hat{R},\hat{\theta})}\left( \begin{array}{c}
    d\hat{R}\\d\hat{\theta}\end{array}\right),
 $$
 and the ECM in the cartesian system is
 \begin{equation}
 \mathbf{\hat{\Sigma}}_\trm{car} = T_\trm{pol}(\hat{R},\hat{\theta})~\mathbf{
 \hat{\Sigma}}_\trm{pol}~T_\trm{pol}(\hat{R},\hat{\theta})^T. \label{eqn:CartCov}
 \end{equation}
 Similarly, for the same receiver at $\underline{X}_c$ receiving prior SLE $\underline{X}_s = [x_s~y_s]^T$ in cartesian coordinates, the prior SLE and ECM can be transformed to the receiver's polar coordinates as
 \begin{eqnarray}
 \bar{R} = ||\underline{X}_c - \underline{X}_s||~\textnormal{and}~ \bar{\theta} = \arg\{\underline{X}_c - \underline{X}_s\},\label{eqn:PolEst}
  \end{eqnarray}
  and
 \begin{equation}
 \mathbf{\hat{\Sigma}}_\trm{pol} = T_\trm{car}(\bar{R},\bar{\theta})~\mathbf{
 \hat{\Sigma}}_\trm{car}~T_\trm{car}(\bar{R},\bar{\theta})^T,\label{eqn:PolCov}
 \end{equation}
 where $T_\trm{car}(R,\theta) = T_\trm{pol}^{-1}(R,\theta)$. Note that the coordinate transformations $T_\trm{pol}$ and $T_\trm{car}$ depend only the measurements and SLE at the current receiver and not on the AoA measurement variance. 

\subsection{Sequential Localization Algorithm}\label{sec:localg}
  We now describe the steps involved in sequentially aggregating receiver AoA estimates to produce a SLE. Let $N$ receivers be located at $\ul{X}^{c}_k = [x_k~y_k]^T$ indexed by $k$, and let the source be at $\ul{X}_s = [x_s~y_s]^T$. Receiver $k$'s AoA estimate is $\hat{\theta}_k$, measured from the x-axis of the global cartesian system. Further, the polar coordinates with receiver $k$ at its origin is designated $\mathcal{P}_k$. For ease of exposition, we assume that the receivers are indexed in the order in which their estimates are combined.
  
  \noindent{\bf The Bootstrap procedure:} The AoA estimates of the first two receivers in the combining order, $\hat{\theta}_1$ and $\hat{\theta}_2$, are used to obtain an initial SLE and ECM to ``bootstrap" the Bayesian algorithm (see Figure \ref{fig:bootstrap}, Left Panel). The estimated range of the source from receiver 1 is
  \begin{equation}
    \hat{R}_1 = \frac{(y_2 - y_1)\cos(\hat{\theta}_2) - (x_2 - x_1)\sin(\hat{\theta}_2)}{\sin(\hat{\theta}_1 -
    \hat{\theta}_2)}.\label{eqn:boots_loc}
  \end{equation}
  Therefore, we can get an initial SLE $\hat{\ul{X}}_s^{(1)} = [\hat{x}_s~\hat{y}_s]^T$ in cartesian coordinates from $\hat{\ul{\mu}}^{(1)} = [\hat{R}_1~\hat{\theta}_1]^T$ using \eqref{eqn:CartEst}. Errors in source range estimates from receivers 1 and 2 can be computed from the errors in the AoA estimates as
  \begin{equation} \label{eqn:boots_cov}
   \left(\begin{array}{c}
    d\hat{R}_1\\d\hat{R}_2\end{array}\right)  = \frac{1}{\sin(\hat{\theta}_1 -
    \hat{\theta}_2)}\left(\begin{array}{cc} -\hat{R}_1\cos(\hat{\theta}_1 -
    \hat{\theta}_2) & \hat{R}_2\\-\hat{R}_1 & \hat{R}_2\cos(\hat{\theta}_1 -
    \hat{\theta}_2) \end{array}\right) \left(\begin{array}{c}
    d\hat{\theta}_1\\d\hat{\theta}_2\end{array}\right).
  \end{equation}
  Using \eqref{eqn:boots_cov} and the conditional independence of the estimates $\hat{\theta}_1$ and $\hat{\theta}_2$ given the source location, the entries of the initial ECM $\Sigma_{RR}$, $\Sigma_{\theta\theta}$, and $\Sigma_{R\theta}$ can be computed in polar coordinates $\mathcal{P}_1$. This initial ECM $\hat{\mathbf \Sigma}_\trm{pol}^{(1)}$ in $\mathcal{P}_1$  is transformed to the global cartesian coordinates as $\hat{\mathbf \Sigma}_\trm{car}^{(1)}$ using \eqref{eqn:CartCov} and $\hat{\ul{\mu}}^{(1)}$. Note that the initial ECM could have been computed in $\mathcal{P}_2$ instead of $\mathcal{P}_1$, but both would ultimately lead to the same $\hat{\mathbf \Sigma}_\trm{car}^{(1)}$. We term these preceding steps the `bootstrap procedure'.

  However, care must be taken in choosing the receivers for the initialization above so as to avoid `bad' initial conditions. It is evident from \eqref{eqn:boots_loc} and \eqref{eqn:boots_cov} that the initial SLE and ECM become unbounded if $(\hat{\theta}_1 - \hat{\theta}_2) \approx n\pi,~n = 0,1,\ldots$, due to the $\sin(\hat{\theta}_1 - \hat{\theta}_2)$ term in the denominator of both equations. Thus, the error is largest when the source is roughly collinear with the two receivers (on the line joining the two receivers), when the nominal AoA measurements are close to being parallel or antiparallel. Therefore, the algorithm is always initialized with a pair of receivers with $\hat{\theta}_1 - \hat{\theta}_2$ significantly different from $0$ or $\pi$, which also ensures that the initial SLE is mostly within the ring of receivers. This idea is similar to the concept of dilution of precision in the global positioning system \cite{GPS} that describes the effect of the satellite configuration on the location accuracy. Thereafter, the receivers can be combined in any random order and effect of this order on performance is simulated in Section \ref{sec:LOSperf}.

  The sequential algorithm for source localization has the following steps:

\textit{Step 1 (Bootstrap):} Estimate initial SLE $\hat{\ul{\mu}}^{(1)}$ and ECM $\hat{\mathbf \Sigma}_\trm{pol}^{(1)}$ in $\mathcal{P}_1$, using \eqref{eqn:boots_loc} and \eqref{eqn:boots_cov}. Transform the SLE and ECM into the global cartesian coordinates as $\hat{\ul{X}}_s^{(1)}$ and $\hat{\mathbf \Sigma}_\trm{car}^{(1)}$, respectively, using \eqref{eqn:CartEst} and \eqref{eqn:CartCov}. Pass $[\hat{\ul{X}}_s^{(1)},\hat{\mathbf \Sigma}_\trm{car}^{(1)}]$ as a \emph{prior} to the next receiver.

\textit{Step 2 (Transformation):} Let the index of the current receiver be $k$. Transform the prior $[\hat{\ul{X}}_s^{(k-1)},\hat{\mathbf \Sigma}_\trm{car}^{(k-1)}]$ to $[\hat{\ul{\mu}}^{(k-1)} ,\hat{\mathbf \Sigma}_\trm{pol}^{(k-1)}]$ in the local polar coordinate system, $\mathcal{P}_{k}$, using \eqref{eqn:PolEst} and \eqref{eqn:PolCov}.

\textit{Step 3 (Aggregation):} Update the prior estimates $[\hat{\ul{\mu}}^{(k-1)} ,\hat{\mathbf \Sigma}_\trm{pol}^{(k-1)}]$ with the AoA estimate of the $k$th receiver, $\hat{\theta}_k$, using the LMMSE procedure in \eqref{eqn:Meanupdate} and \eqref{eqn:Covupdate}. Transform updated estimates $[\hat{\ul{\mu}}^{(k)},\hat{\mathbf \Sigma}_\trm{pol}^{(k)}]$ into the global cartesian coordinates as $[\hat{\ul{X}}_s^{(k)},\hat{\mathbf  \Sigma}_\trm{car}^{(k)}]$.

\textit{Step 4 (Termination):} If there are unprocessed AoA measurements, pass priors on to the next unaggregated receiver and go to \textit{Step 2}. Otherwise output the SLE and stop.

Since only the current SLE and ECM need to be passed on from one receiver to the next, this algorithm can be implemented in a distributed manner, with each receiver needing to know only its own location and orientation. While we consider AoA estimates here, this sequential algorithm is quite general, and can, for example, incorporate probabilistic information on the source range obtained from signal strength measurements. Further, the algorithm is scalable, in that its complexity grows only linearly in the number of receivers. This scalability is required to realize the improvement in localization performance with the number of receivers, details of which are given in Section \ref{sec:CRLB}.

\subsection{ML Estimator and Cramer-Rao Bound}\label{sec:CRLB}
The CRLB is a lower bound on the localization performance of the best minimum variance unbiased estimator. For analytical simplicity, we work with the Gaussian AoA model in \eqref{eqn:LOSmodel}, although corresponding bounds for other models such as the Laplacian in \eqref{eqn:Lapmodel} can be easily derived. For the Gaussian AoA error model, the log-likelihood function (within scale factors and constants) for the observed AoA estimates $\{\hat{\theta_i}\}_{i=1}^N$ given the source location $\ul{X} = (x,y)$ is
\begin{equation}\label{eqn:gausloc}                                                                                      L(\hat{\theta}_1,\hat{\theta}_2,\ldots,\hat{\theta}_N/\ul{X}) = -\sum_{k=1}^{N} \frac{(\hat{\theta}_k - \theta_k(\ul{X}))^2}{\sigma_k^2},
\end{equation}
where $\theta_k(\ul{X})$ is the true bearing of the source from receiver $k$ and $\sigma^2$ is the AoA spread (estimation error variance) at receiver $k$. The ML estimator searches for the location $\ul{X}$ that maximizes the log-likelihood function in \eqref{eqn:gausloc}, which is a nonlinear least squares problem. For small $\sigma_k^2$, the cost function in \eqref{eqn:gausloc} can be shown to be approximately concave. We have observed numerically that the cost function has an unique global maxima for the parameter values of interest and a standard nonlinear least squares solver (such as the ``lsqnonlin'' function in Matlab$^\textnormal{\textregistered}$ with default parameters that is based on \cite{Coleman1996}) produces the ML estimate. The ML estimate is shown to achieve the CRLB in Section \ref{sec:LOSperf}.
\begin{figure*}[ht]%
\centering
\subfloat[][]{\includegraphics[width=3.4in,clip]{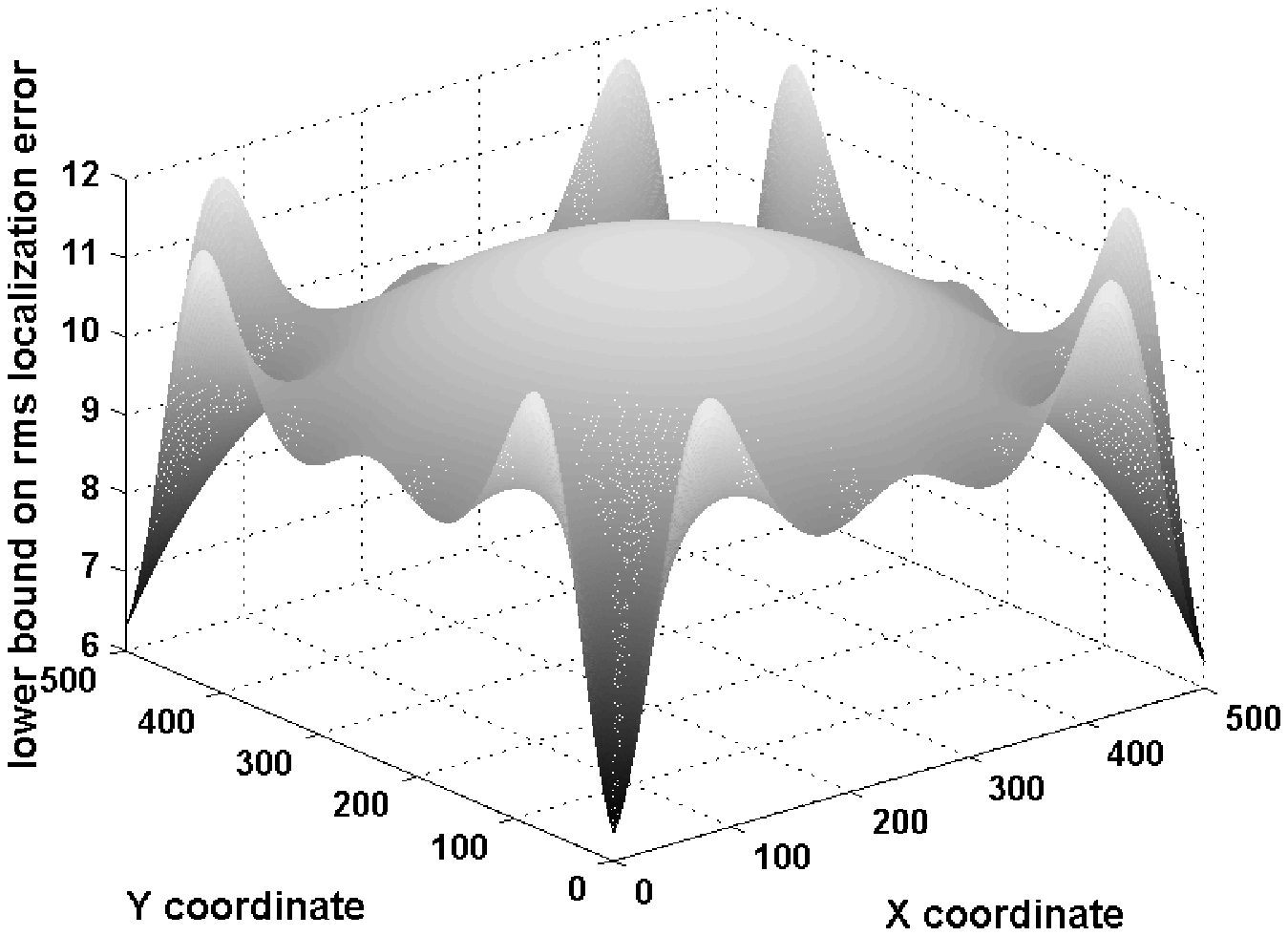}}%
\subfloat[][]{\includegraphics[width=3.4in,clip]{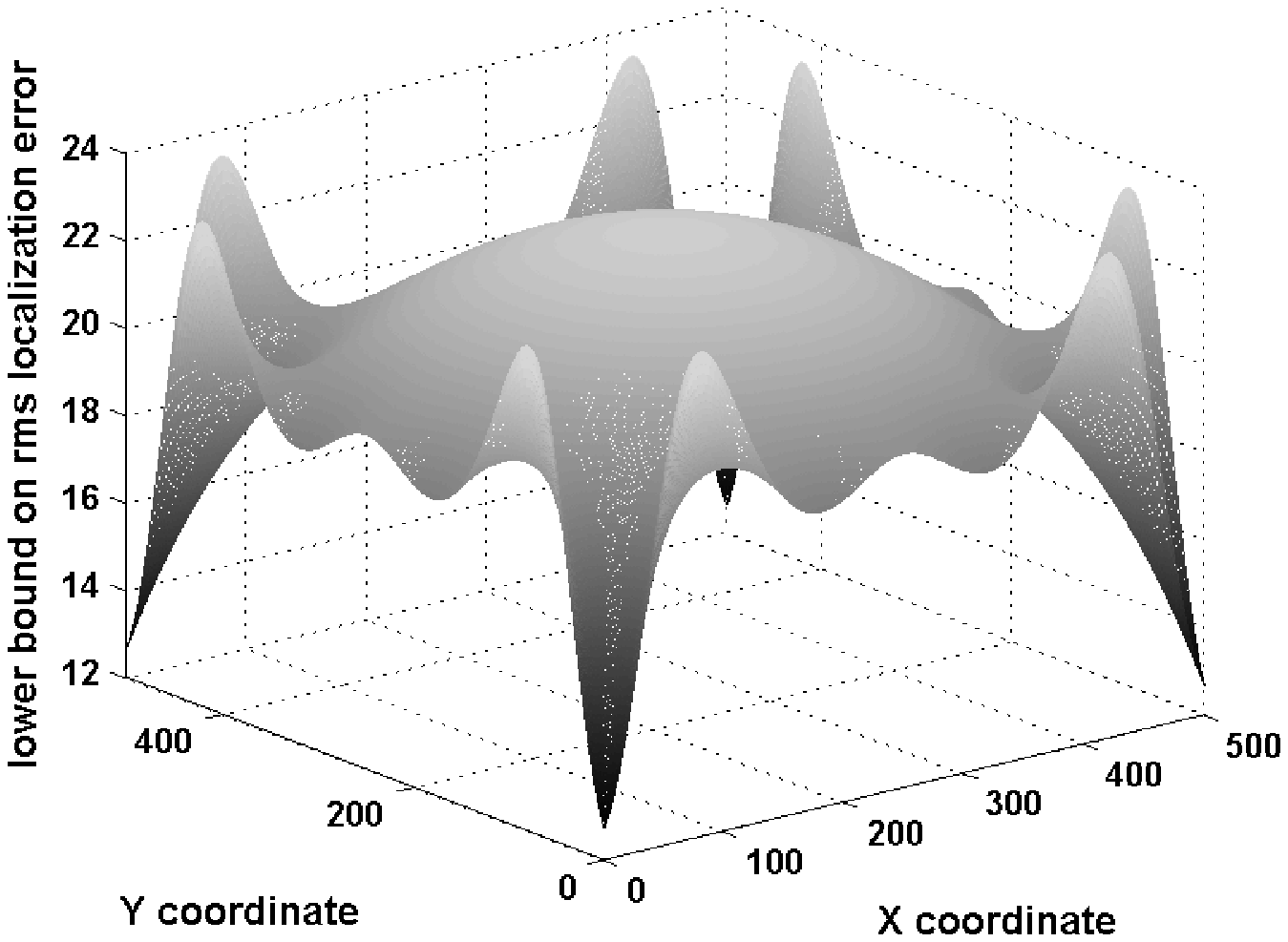}}%
\caption{Surface plots of the CRLB at various locations on a 500x500 square sensor field surrounded by 8 receivers in a circle for $2^\circ$ (Left) and $4^\circ$ (Right) AoA spread are shown. The collectors are at $(250 + 457\cos\{2\pi (k-1)/8\},250 + 457\sin\{2\pi (k-1)/8\}),~k=1,...,8$. The CRLB is highest at the center as predicted but some regions close to the receivers also have high bounds. The CRLB contours display multiple symmetries due to the equispaced and circular geometry of receivers.}%
\label{fig:CRLBvar}%
\end{figure*}
 
We first construct the Fisher information matrix (FIM) ${\mathbf J}(\ul{X})$ that is the inverse of the CRLB ${\mathbf \Sigma}_\textrm{CR}(\ul{X}) = {\mathbf J^{-1}(\ul{X})}$. For the Gaussian LOS model in \eqref{eqn:LOSmodel}, 
\begin{equation}                                                                                                         {\mathbf J}(\ul{X}) =  \left(\begin{array}{cc}
\sum_{k=1}^N \frac{\sin^2(\theta_k)}{\sigma_k^2 R_k^2} & -\sum_{k=1}^N \frac{\cos(\theta_k)\sin(\theta_k)}{\sigma_k^2 R_k^2} \\
-\sum_{k=1}^N \frac{\cos(\theta_k)\sin(\theta_k)}{\sigma_k^2 R_k^2} & \sum_{k=1}^N \frac{\cos^2(\theta_k)}{\sigma_k^2 R_k^2} \\
\end{array}\right),\label{eqn:FIM}
\end{equation}
where $\ul{\hat{\theta}} = (\hat{\theta}_1,\hat{\theta}_2,\ldots,\hat{\theta}_N)$, and $R_k$ and $\theta_k$ are the range and bearing of the source at location $\ul{X}$ measured from receiver $k$. The total localization error variance is the sum of the variances along the $x$ and $y$ coordinates. The lower bound on the total localization error is
\begin{equation} \label{eqn:CRLB}
    \sigma_\trm{tot}^2(\ul{X})~\geq ~\trm{Tr}({\mathbf \Sigma}_\textrm{CR}(\ul{X})) = \frac{\sum_{k=1}^{N}\frac{1}{R_k^2\sigma_k^2}}{\sum_{k=1}^{N}\sum_{l=k+1}^{N}\frac{\sin^2(\theta_k - \theta_l)}{R_l^2R_k^2\sigma_k^2\sigma_l^2}}.
\end{equation}
Observe that the CRLB is dependent on the location of the source $\ul{X}$ through $\{R_k\}_{k=1}^N,\{\theta_k\}_{k=1}^N$ (see Figure \ref{fig:CRLBvar}). To gain insight into the factors determining the localization error, consider an example of $N$ receivers equally spaced on the perimeter of a disc of radius $R$ (as in Figure \ref{fig:bootstrap}). We simplify the analysis further by computing the bound in \eqref{eqn:CRLB} for a source at the center of the disc, with the assumption that the spatial spread in the AoA is the same for all receivers (we describe a way of handling different AoA spreads at each receiver in Section \ref{sec:aggr_outlierdet}). It can be verified numerically that the CRLB is maximized at the center of disc (see Figure \ref{fig:CRLBvar}), and therefore corresponds to the worst source position from the standpoint of localization performance. By realizing that $\theta_k = 2\pi k/N$ for a source at the center of the disc, the lower bound on the localization error at the center of the disc can be computed as
\begin{equation} \label{eqn:centerbound}
    \sigma_\trm{tot}^2~\geq ~ \frac{NR^2\sigma^2}{\sum_{k=1}^{N}\sum_{l=k+1}^{N}\sin^2(\theta_k - \theta_l)}\approx \frac{2R^2\sigma^2}{\pi^2(N-1)},
\end{equation}
using the fact that $\sin(\theta)\leq 1$. 
Even with a conservative bound, we observe that the localization error reduces at least inversely with the number of receivers, thus providing a method of reducing the localization error by increasing the number of receivers. The localization error grows linearly with the AoA spread. As seen in Figure \ref{fig:CRLBvar}, although the localization error is maximum at the center of the sensor field, the variation of localization error is more complex close to the edges of the field. Moreover, AoA measurements are less accurate for far away sources based on \eqref{eqn:centerbound} and discussion on bootstrapping. Thus, while we consider a ring of receivers for simplicity, the optimal placement in order to minimize the worst-case localization error (assuming a fixed area to be covered by a fixed number of receivers) is an interesting open issue: we wish to reduce the worst-case distance of a source from the receivers nearest it, but we also wish to have a sufficient number of receivers that are near enough to a source.

\subsection{Properties of the Sequential Algorithm}\label{sec:props}
  The ML source location is the solution to the non-linear least squares (LS) problem in \eqref{eqn:gausloc}. The solution to this LS problem is asymptotically (in the number of receivers $N$) consistent and efficient, when the observation noise is Gaussian (e.g., LOS AoA error model in \eqref{eqn:LOSmodel}), and the LS solution is asymptotically consistent even when the noise is non-Gaussian, but with no guarantees on efficiency \cite{poor_book}. The ML estimate can be computed maintaining the asymptotic consistency and efficiency with the following recursive algorithm \cite{poor_book}:
  \begin{equation}\label{eqn:MLrecurse}
     \hat{\ul{X}}_n = \hat{\ul{X}}_{n-1} + {\mathbf J}^{-1}(\hat{\ul{X}}_{n-1})\frac{\partial L(\hat{\theta}_n/\hat{\ul{X}}_{n-1})}{\partial \ul{X}},~~n = 2,...,N,
  \end{equation}
where $\hat{\ul{X}}_n$ is the source location after combining the first $n$ receiver AoA estimates, ${\mathbf J}(\hat{\ul{X}}_{n-1})$ is the FIM for localization with only the first $n-1$ receivers, and $L(\hat{\theta}_n/\hat{\ul{X}}_{n-1})$ is the contribution of the $n$th observation to the log-likelihood function in \eqref{eqn:gausloc}. As in Section \ref{sec:CRLB}, let $[R_n~ \theta_n]^T$ represent the current SLE $\hat{X}_{n-1}$ in the polar coordinates centered at receiver $n$. We can now rewrite the recursion for this localization problem as
  \begin{equation}
     \hat{\ul{X}}_n = \hat{\ul{X}}_{n-1} + {\mathbf J}^{-1}(\hat{\ul{X}}_{n-1})\left(
                                                                                 \begin{array}{c}
                                                                                   R_n\sin(\theta_n) \\
                                                                                   -R_n\cos(\theta_n) \\
                                                                                 \end{array}
                                                                               \right)(\hat{\theta}_n - \theta_{n}(\hat{X}_{n-1})),~~n = 2,...,N.\nonumber
  \end{equation}
  The above equation has exactly the same form as the LMMSE update in \eqref{eqn:linupdate}, \eqref{eqn:kalgain}, \eqref{eqn:Meanupdate} with the only difference being that this is an update in the global cartesian coordinates rather than the local polar coordinates of receiver $n$. The Hessian-like FIM ${\mathbf J}$ is also updated at each step and new FIM ${\mathbf J}(\ul{X}_n)$ can be computed using \eqref{eqn:FIM}:
  \begin{eqnarray}
      {\mathbf J}(\hat{\ul{X}}_n) &=& {\mathbf J}(\hat{\ul{X}}_{n-1}) + \left(
                                                                        \begin{array}{cc}
                                                                          \frac{\sin^2(\theta_n)}{\sigma_n^2 R_n^2} & \frac{-\cos(\theta_n)\sin(\theta_n)}{\sigma_n^2 R_n^2}\\
                                                                          \frac{-\cos(\theta_n)\sin(\theta_n)}{\sigma_n^2 R_n^2} & \frac{\cos^2(\theta_n)}{\sigma_n^2 R_n^2} \\
                                                                        \end{array}
                                                                      \right)\nonumber\\
                                  &=& {\mathbf J}(\hat{\ul{X}}_{n-1}) + (T_\trm{pol}^{-1}(R_n,\theta_n))^H\left(
                                                                        \begin{array}{cc}
                                                                           0 & 0\\
                                                                           0 & \frac{1}{\sigma_n^2}
                                                                        \end{array}
                                                                       \right)T_\trm{pol}^{-1}(R_n,\theta_n)\label{eqn:recur_cov_upd},
  \end{eqnarray}
  where $T_\trm{pol}(R_n,\theta_n)$ is the unitary matrix, in \eqref{eqn:CartCov}, used to convert ECM from local polar coordinates to global cartesian coordinates. Recalling that the FIM is the inverse of the covariance matrix, observe that the covariance update in \eqref{eqn:recur_cov_upd} is equivalent to the update in the sequential algorithm in \eqref{eqn:cov_mat_update} with appropriate transformations to cartesian coordinates. Thus, the sequential algorithm is an alternate formulation of this recursive ML computation, and therefore inherits the properties of the recursive ML algorithm. However, LS solutions are sensitive to \emph{outliers} in the observations and we present a robust extension of this algorithm next in Section \ref{sec:NLOSalg}.

\section{Localization with Non-line-of-sight Channels}\label{sec:NLOSalg}

We employ the LOS and NLOS models described in Section \ref{sec:models} with AOA estimates corresponding to either strong multipath components or those that are modeled as \emph{outliers}, which are chosen uniformly from the set of feasible angles. We present an algorithm capable of \emph{outlier} suppression to handle both the narrowband and wideband scenarios. However, we focus on the former (where each receiver produces a single AoA estimate) due to its simplicity and later indicate how the algorithm can incorporate multiple AoA estimates at each receiver. 

\subsection{ML Localization under NLOS propagation}\label{sec:MLmethod}
Let $\alpha$ be the fraction of receivers that experience stronger contributions to the received signal from the NLOS components than the LOS components. Under the model in \eqref{eqn:narrowbndmod}, the log-likelihood function for the source location $\ul{X}$, where receiver $k$ has AoA spread $\sigma_k$, is 
\begin{equation} \label{eqn:nonLOSmod}
  L(\ul{\hat{\theta}}/\ul{X}) = \sum_{k=1}^{N} \log \left[\frac{1-\alpha}{\sigma_k\sqrt{2\pi}(1-2Q(\frac{\pi}{2\sigma_k}))}\exp\left(-\frac{(\hat{\theta}_k - \theta_k(\ul{X}))^2}{2\sigma_k^2}\right) + \frac{\alpha}{\pi} \right],
\end{equation}
where $\ul{\hat{\theta}}$ is the vector of observed AoAs. Equation \eqref{eqn:nonLOSmod} can be approximated using $\log(e^a + e^b)\approx \min(a,b)$ as
\begin{equation} \label{eqn:approxllf}
  L(\ul{\hat{\theta}}/\ul{X}) \approx -\sum_{k=1}^{N} \min \left[(\hat{\theta}_k - \theta_k(\ul{X}))^2,\Theta_{\trm{max},k}^2 \right],
\end{equation}
where
\begin{equation}\label{eqn:thetamax}
\Theta_{\trm{max},k}^2 = 2\sigma_k^2 \log\left(\frac{\sqrt{\pi} (1-\alpha)}{\sigma_k\sqrt{2}\alpha(1-2Q(\frac{\pi}{2\sigma_k}))}\right).
\end{equation}
Then the ML SLE $\hat{\ul{X}}$ is
\begin{equation} \label{eqn:MLest}
  \hat{\ul{X}} = \arg\max_{\ul{X}}~ L(\ul{\hat{\theta}}/\ul{X}) \approx \arg\min_{\ul{X}}~\sum_{k=1}^{N} \min \left[(\hat{\theta}_k - \theta_k(\ul{X}))^2,\Theta_{\trm{max},k}^2 \right].
\end{equation}
The ML algorithm in \eqref{eqn:MLest} is a standard LS minimization with angular errors bounded by a threshold $\Theta_{\trm{max},k}$ for receiver $k$. This cost function ensures that there is no incentive to reducing angular errors larger than the threshold and therefore, outliers or NLOS estimates only have a limited effect on the SLE. As desired, the ML estimator in \eqref{eqn:MLest} reduces to the ML method for the LOS scenario (see \eqref{eqn:gausloc}) in the absence of outliers ($\alpha \rightarrow 0$), and the threshold $\Theta_{\trm{max},k} \rightarrow \infty$. Although the this ML algorithm is computationally prohibitive, we present it to provide a heuristic for the \emph{outlier} suppression algorithm.

For simplicity of exposition, we describe the outlier suppression algorithm with the same AoA spread for all receivers and hence, a common threshold $\Theta_\trm{max}$. To extend the sequential algorithm to the NLOS scenario, we impose this constraint, $\Theta_\trm{max}$, on the largest observed angular error at each step and describe this modified algorithm in the following section.

\subsection{Sequential Aggregation with Outlier Suppression}\label{sec:aggr_outlierdet}
 In NLOS scenarios, the localization algorithm must identify the largest subset of receivers with LOS channels that are mutually consistent, and use these AoA measurements to estimate the source location. However, the receivers with NLOS channels (outliers) are unknown and arbitrary in number. This search for the largest mutually consistent set of receivers is of exponential complexity, since it must explore every subset of the $N$ receivers. We therefore resort to a randomization of the algorithm in Section \ref{sec:localg}, in which we randomize the choice of the first two receivers used in the bootstrap phase. Thereafter, each subsequent receiver's AoA estimate is combined ensuring that angular errors in all the aggregated receivers remain below the threshold $\Theta_\trm{max}$ from \eqref{eqn:thetamax} with $\sigma_k = \sigma$. By bootstrapping with different pairs of receivers, this randomized algorithm produces source position estimates corresponding to different subsets of receivers that mutually agree, leading to a list of possible explanations for the observed $\hat{\ul{\theta}}$.

 The outlier suppression algorithm is an extension of the sequential algorithm in Section \ref{sec:localg} with two key differences: First, at each step, the next receiver chosen for aggregation is the one with the smallest angular error. The angular error at a receiver is the discrepancy between the bearing of a hypothetical source at the current estimated location, $\hat{\ul{X}}$, and the AoA estimate measured at that receiver, i.e., $e(\hat{\theta},\hat{\ul{X}}) = |\hat{\theta}-\theta (\hat{\ul{X}})|$. The error $e$ can be understood as the empirical estimate of the error in \eqref{eqn:MLest}.  The receiver with the smallest angular error is the receiver whose AoA estimate is most consistent with the current SLE. Second, after combining a new AoA estimate, the updated source location is retained only if the angular errors for all the aggregated receivers are below the threshold $\Theta_\trm{max}$. This ensures that outliers are eliminated using the criterion in \eqref{eqn:MLest} and simultaneously, the receivers with LOS are used to compute the source location using an LS computation. 
 
Sequential aggregation with outlier suppression involves repetitions of the following basic steps with multiple random bootstraps (for brevity, we use the phrase ``combine with $\hat{\theta}$" to denote the computation of the new SLE and ECM using the appropriate coordinate transformations described in Section \ref{sec:comb_tform}):

 \textit{Step 0 (Initialization):} Set the list of receivers already aggregated $\mathcal{A} = \emptyset$ and list of receivers yet to be combined, $\mathcal{C} = \{1,\ldots,N\}$.

 \textit{Step 1 (Bootstrap):} Select a pair of receivers at random, say $\{i,j\}$ that has not been used previously. Compute the initial estimate $\hat{\ul{X}}$ and ECM $\hat{\mathbf \Sigma}$ using the bootstrap procedure. Add $\{i,j\}$  to the list of aggregated (inlying) receivers, $\mathcal{A} = \mathcal{A}\cup \{i,j\}$, and remove it from the list of remaining receivers $\mathcal{C} = \mathcal{C}- \{i,j\}$.

 \textit{Step 2 (Angular Error Computation):} Compute angular errors $e(\hat{\theta},\hat{\ul{X}})$ over the remaining receivers $\mathcal{C}$, i.e.,
 $$
     e(\hat{\theta}_i,\hat{\ul{X}}) = |\hat{\theta}_i-\theta_i (\hat{\ul{X}})|~~\forall~i\in \mathcal{C} .
 $$

 \textit{Step 3 (Candidate Selection and Aggregation):} Find the receiver $k$ with the smallest error $e(\hat{\theta}_k,\hat{\ul{X}})$. Combine $\hat{\ul{X}}$ with $\hat{\theta}_k$ to obtain the new \emph{candidate} estimate $\hat{\ul{X}}{'}$ and covariance $\hat{\mathbf \Sigma}{'}$.

 \textit{Step 4 (Threshold Verification):} If the angular error using the \emph{candidate} location $\hat{\ul{X}}{'}$ is below the threshold for all the aggregated receivers, i.e.,
$$
  e(\hat{\theta}_l,\hat{\ul{X}}{'})<\Theta_\trm{max}, ~\forall~l \in \mathcal{A}\cup \{k\},
$$
 then retain the \emph{candidate} location and covariance,  $\hat{\ul{X}} = \hat{\ul{X}}{'}$, and $\hat{\mathbf \Sigma}= \hat{\mathbf \Sigma}{'}$. Also add receiver $k$ to the list of aggregated receivers, $\mathcal{A} = \mathcal{A}\cup \{k\}$.

 \textit{Step 5 (Termination):} Remove receiver $k$ from further consideration $\mathcal{C} = \mathcal{C} - \{k\}$. If there are no remaining receivers ($\mathcal{C} = \emptyset$) then Stop else goto \textit{Step 2}.

Here we assumed that the AoA spread and hence, the threshold is the same for all receivers. As seen from \eqref{eqn:MLest}, the error at each receiver $k$ is only compared against $\Theta_{\trm{max},k}$, the threshold dependent on AoA spread for receiver $k$. Thus, the above NLOS suppression algorithm can be applied to different AoA spreads by replacing the threshold in Step 4 by the appropriate $\Theta_{\trm{max},k}$ from \eqref{eqn:thetamax}.

The above algorithm is repeated with $M$ different random initial conditions in order to detect the source with a high probability, and each run produces a likely source location and a confidence (the ECM) in that estimate. In a wideband system, if a receiver resolves two arriving paths, the above algorithm can still be used by introducing a second \emph{virtual} receiver at the same location as the original receiver and assigning to it the second arriving path. However, it must be ensured that both the original and virtual receivers are not part of the same SLE, since we cannot have two different AoA estimates at a given receiver corresponding to an LOS path. This situation is illustrated with an example in Section \ref{sec:NLOSperf}.

{{\bf Choice of} $M$ {\bf and} $\Theta_\trm{max}${\bf :}}
The performance of the outlier suppression algorithm is determined by the choice of the maximum angular error $\Theta_\trm{max}$ and the total number of iterations, $M$. Assuming that the spread in AoA is known at each receiver, the threshold $\Theta_\trm{max}$ can be computed using \eqref{eqn:thetamax}, if the fraction of receivers with strong NLOS components $\alpha$ is known. In practice, the largest expected fraction of receivers with strong NLOS components, $\alpha_\trm{max}$, can be set based on knowledge of the propagation environment and the worst case NLOS scenario under which we wish to operate. This value of $\Theta_\trm{max}$ obtained using \eqref{eqn:thetamax} is conservative for lower levels of multipath scattering, as $\Theta_\trm{max}$ is monotonically decreasing with $\alpha$. While choosing a smaller $\Theta_\trm{max}$ might prevent some ``good'' LOS AoA estimates from being utilized, it also ensures that NLOS measurements do not corrupt the SLE.

During the bootstrap phase, two receivers are chosen randomly to seed the sequential algorithm. Success in the sequential estimation depends on selecting two receivers with LOS channels to initiate the algorithm. We have observed from simulations that the algorithm always converges to a solution in the `vicinity' of the bootstrap location, and therefore an estimate using almost all the LOS estimates is produced, if the algorithm is bootstrapped with two receivers having LOS channels. Hence, we hypothesize that bootstrap failure is the predominant cause of localization failure (we verify this numerically in Section \ref{sec:NLOSperf}). We now try to estimate this probability of failure as a function of $M$, the number of iterations of the outlier suppression algorithm. Suppose the outlier suppression algorithm is repeated with $M$ \emph{different} seeds, the probability of failure in the bootstrap phase is the probability that at least one receiver is an outlier in each of the $M$ attempts. The total number of bootstrap pairs is $P = \binom{N}{2}$ and number of pairs with at least one NLOS receiver is $K = P - \binom{\lfloor (1-\alpha) N\rfloor}{2}$, where $\lfloor (1-\alpha) N\rfloor$ is the number of receivers with LOS channels. Then,
\begin{equation}\label{eqn:boot_failure}
   P(\trm{bootstrap failure}) = \left\{\begin{array}{ll}
                  \frac{\binom{M}{K}}{\binom{P}{M}} & \trm{if }~ M\leq K\\ 0 & \trm{if }~ M>K
                  \end{array}\right.
\end{equation}
When the receivers resolve multiple arriving paths, the above probability of failure computation is modified by replacing the number of receivers by the total number of resolved paths. Depending on the probability of failure acceptable in the system, \eqref{eqn:boot_failure} is used to choose the number of randomizations of the algorithm that are necessary.

 A practical issue of interest is the choice of the number of randomizations $M$ for different total number of receivers $N$ to achieve a fixed probability of bootstrap failure under similar NLOS propagation environments (fixed $\alpha$). Rearranging the tight upper bound on the probability of bootstrap failure, presented in the Appendix as \eqref{eqn:bootupbnd}, we get an tight upper bound on the possible value of $M$ as
 \begin{equation}
    M\leq \frac{\log (P(\trm{bootstrap failure}))}{\log (1-(1-\alpha)^2)},
    \label{eqn:boot_bound}
 \end{equation}
 using the fact that for large N
 $$
   \frac{K}{P} = 1 - \frac{\binom{\lfloor (1-\alpha) N\rfloor}{2}}{\binom{N}{2}}\approx 1-(1-\alpha)^2.
 $$
We observe that the probability of bootstrap failure is independent of $N$. Thus, the outlier suppression algorithm with complexity $O(MN^2)$ is still only quadratic in the number of measurements.
 
\section{Numerical Results}\label{sec:sims}
  We study the performance of the proposed algorithms via Monte-Carlo simulations. The simulation setup is as follows: We consider a circular field of unit radius with $N$ equally spaced receivers along the perimeter. As seen from our analysis in \eqref{eqn:centerbound}, the localization error grows linearly with distance from the receiver. Therefore, by selecting a field of unit radius, we obtain scale-invariant (only dependent on dimensionless quantities) measures of performance. Let the receivers be located at $[\cos(2\pi(k-1)/N)~\sin(2\pi(k-1)/N)]^T, k = 1,\ldots,N$. Each receiver measures the AoA of the signal received at its antenna array and representative AoA estimates are generated according to the models described in Section \ref{sec:models}. We assume throughout that all the receivers have the same AoA spread for convenience, although the algorithm does not require this, and that only a single source transmits at any given time. 
    
  The AoA spreads under different environments reported in literature are listed in Table \ref{tab:aoaspreads}. The effective AoA spread for our NLOS models in \eqref{eqn:narrowbndmod} is the quantity reported in most studies. This effective spread is greater than the AoA spread $\sigma$ for the Gaussian LOS model alone due to the mixture with the uniform distribution corresponding to LOS blockage. Therefore, accounting for the available information on propagation environments in Table \ref{tab:aoaspreads}, we consider AoA spreads $\sigma$ in the range $1 - 10^\circ$ for numerical studies of our algorithm. We use the CRLB \eqref{eqn:CRLB} as a performance benchmark, but this bound is dependent on the true location of the source. In order to have a fair comparison, we run equal number of iterations on each of 25 candidate source locations, and compare the total rms localization error against the average of the CRLB at those 25 locations.

\begin{table}[ht]
	\centering
	  \caption{AoA spreads from reported measurements in different environments}\label{tab:aoaspreads}
		\begin{tabular}{|c|c|c|}\hline
		 Environment & AoA spread & Reference\\\hline
		 Outdoor - Urban & $\sim 10^\circ $ (LOS + NLOS) &  Figures 2 and 3, Klein et al. \cite{Klein1996}\\\hline
		 Outdoor - Urban & $\sim 15^\circ$ (LOS + NLOS)& Figure 5, Thomas et al. \cite{Thomas1992}\\\hline
		 Indoor - LOS blockage & $20 - 30^\circ$ (NLOS)& Spencer et al. \cite{Spencer1997} \\\hline
		 Indoor - WIMAX  & $10^\circ$ (LOS), $>50^\circ$ (NLOS) & Figures 6 and 7, Akdhar et al. \cite{Akhdar2009}\\\hline
		 Outdoor - Urban & $3-20^\circ$ (LOS) depending on distance & Figure 10, Chen and Asplund	\cite{Chen2001}\\\hline
		\end{tabular}
\end{table}
%\begin{figure}[h!]
% \centering
%    \includegraphics[width=3.5in]{LOSlocerr}%
%    \caption{Localization performance of the sequential estimation algorithm (in dashed lines) in the LOS scenario for $N = 6,8,12$ receivers. The performance is compared against the CRLB (in solid lines) and the ML estimator (in dotted lines).}\label{fig:LOSlocerr}%
%\end{figure}
\subsection{Performance under LOS scenarios}\label{sec:LOSperf}
\begin{figure}[h!]
\centering
 \includegraphics[width=3.5in]{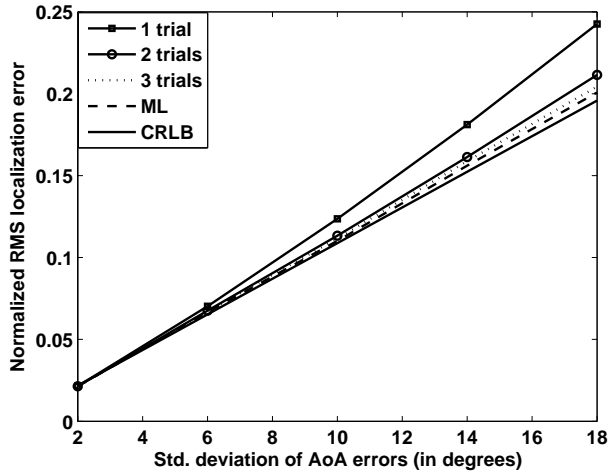}%
  \caption{Performance of sequential algorithm under LOS setting for $N=8$ receivers is compared against the CRLB (solid) and the ML estimate (dashed). The performance improvement using multiple random bootstraps (1,2 or 3) is also shown.}\label{fig:LOSimprove}%
\end{figure}

  The performance of the sequential algorithm in Section \ref{sec:localg} for $N=8$ receivers is shown in Figure \ref{fig:LOSimprove}. The algorithm achieves the CRLB for small angular estimation errors (in simulations for as few as 6 receivers), but the performance deteriorates for large spatial spreads in AoA. The optimal ML estimator, in Section \ref{sec:CRLB}, can be shown to be approximately convex, and the likelihood function has a unique maxima. The sequential algorithm, which is an instance of a stochastic approximation algorithm \cite{poor_book}, has a tendency to get stuck in local minima that arise with large AoA spreads; at large spreads, there are multiple locations where subsets of estimated AoAs are (roughly) concurrent, but no single location close to all the directions is evident. 
  
  This problem is solved by selecting the most likely SLE (i.e., has the highest ML cost in \eqref{eqn:gausloc}) from multiple runs of the sequential estimation using different pairs of receivers to bootstrap. From the simulations in this section, it appears that the choice of a random combining order for the AoA measurements does not incur any significant loss in performance against the CRLB. In reality, the nonlinear coordinate transformations at each step in the sequential algorithm are dependent on the current SLE making the final estimate weakly dependent on the specific order. Nevertheless, the remaining gap to the ML performance can be closed by using multiple random bootstraps and selecting the estimate with smallest ML cost, as shown in Figure \ref{fig:LOSimprove}, and conforms well with the randomization framework in the NLOS suppression algorithm in Section \ref{sec:aggr_outlierdet}.
\begin{figure}[ht]
\centering
\includegraphics[width=3.5in]{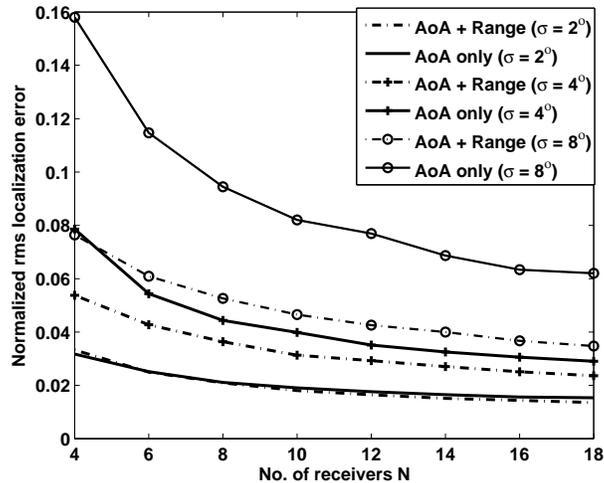}%
 \caption{Comparison of localization performance of the sequential algorithm in a system with only AoA information and one where all nodes have RSS-based range estimates in addition to AoA measurements for different numbers of receivers. The receiver range estimate is modeled as a log-normal random variable parameterized by SNR = $12$ dB.}%
 \label{fig:LOSrangeinfo}%
\end{figure}  

 {\noindent \bf Incorporating range information:} The localization resolution can also be improved by using additional modalities such as received signal strength (RSS)-based ranging. In order to demonstrate the flexibility of our algorithm to also leverage range information, we compare our system, utilizing only AoA measurements, to one where all the receivers additionally obtained RSS-based range information. The RSS-based range estimate, $\tilde{R}$, is modeled as a log-normal random variable (independent of the AoA estimate) as suggested in (7) within \cite{Patwari2003}: $\tilde{R} = R\exp{(\frac{r}{\beta \textrm{SNR}}})$, where $R$ is the true source range, $r$ is zero-mean unit variance Gaussian random variable, and $\beta$ is the path loss exponent. For the simulations, we chose $\beta = 3$ and the variance of the random variable $\tilde{R}$ was taken to be the measurement error variance. The same update formulation in Section \ref{sec:comb_tform} can be extended to include range information as well. Given the new range and AoA estimates $\ul{\tilde{\mu}} = [\tilde{R}~~\tilde{\theta}]^T$ with covariance $\tilde{\mathbf\Sigma}$, the Kalman gain $\mathbf{K}$ is computed as $\mathbf{K} = \mathbf{\bar{\Sigma}}(\mathbf{\bar{\Sigma}} + \tilde{\mathbf\Sigma})^{-1}$ and thus, the new SLE is $\ul{\hat{\mu}} = \ul{\bar{\mu}} + \mathbf{K}(\ul{\tilde{\mu}} - \ul{\bar{\mu}})$. The updated covariance matrix is ${\mathbf{\hat{\Sigma}}}^{-1} = \mathbf{\bar{\Sigma}}^{-1} + \tilde{\mathbf{\Sigma}}^{-1}$. 
 
 In Figure \ref{fig:LOSrangeinfo}, we show the normalized localization performance for three scenarios where the percentage errors in the AoA measurement are below (bottom curves), approximately equal (middle curves) and greater (top curves) than the percentage errors in the range estimates, respectively. As expected, as the quality of the range information improves, the gains from using range information are greater. Further, when the range measurements are less accurate than the AoA estimates, performance with hybrid range and AoA information can be approached using AoA measurements only, with moderate increases in the the number of receivers, $N$. This is possible as location errors due to both RSS-based range estimates and AoA estimates scale linearly with the range of the source from the receiver, allowing the substitution of range information by AoA estimates from additional receivers. Moreover, this similarity in scaling makes the improvement in localization resolution with number of receivers independent of the modality used (see Figure \ref{fig:LOSrangeinfo}). Finally, a key attribute that we reiterate here is the capability of this framework to easily incorporate other sensing modalities. Although we do not explicitly demonstrate here, outliers in the range estimates can be handled similarly in the algorithm in Section \ref{sec:aggr_outlierdet} by adding a range threshold.

\subsection{Performance under NLOS scenarios}\label{sec:NLOSperf}
 In this section, we explore the capabilities of the outlier suppression algorithm under the narrowband and wideband multipath models.
\begin{figure}[ht]
\centering
 \includegraphics[width=3.5in]{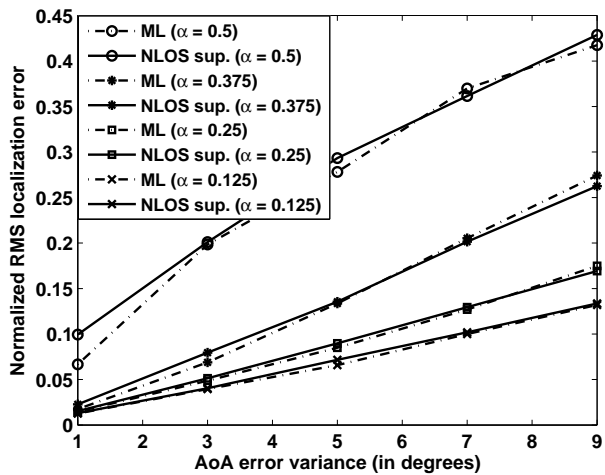}%
  \caption{Localization performance with NLOS suppression with $N = 8$ receivers for different fractions, $\alpha$, of receivers with NLOS channels with optimal thresholds chosen according to \eqref{eqn:thetamax}. The location error variance (solid line) is compared against the ML error variance (dash-dotted line).}\label{fig:NLOSlocerr}%
\end{figure}

\subsubsection{Effect of multipath in a narrowband system}
  We first present numerical results for the narrowband multipath model, where each receiver can only resolve paths spatially and each receiver produces a single AoA estimate corresponding to either the LOS path, or the reflected and scattered multipath. In Figure \ref{fig:NLOSlocerr}, we compare the localization error using the outlier suppression algorithm and the optimal ML estimator in Section \ref{sec:MLmethod} for a system with 8 receivers. For different fractions of outliers, $\alpha$, the outlier suppression algorithm is run with the threshold $\Theta_\trm{max}$ chosen according to \eqref{eqn:thetamax}, while the number of random bootstraps is chosen to ensure that the probability of bootstrap failure is less than $10^{-3}$. This resulted in a choice of the number of random seeds, $M = \{4,7,11,15\}$, for $\alpha = \{0.125,0.25,0.375,0.5\}$ (or $\{1,2,3,4\}$ outliers) using \eqref{eqn:boot_bound}. The algorithm puts out multiple solutions, one corresponding to each random initialization. After pruning out the estimates that placed the source outside the circular field, the ML cost function in \eqref{eqn:nonLOSmod} is used to select the most likely estimate. The ML estimate was obtained by brute force minimization of the same cost function. 
  
\begin{figure}[ht]
\centering
\includegraphics[width=3.5in]{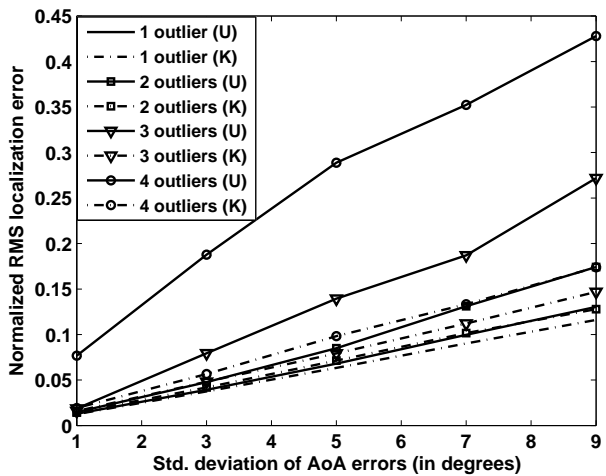}%
\caption{The ML localization performance for an $N = 8$ receiver system with different numbers of receivers with NLOS channels for the two situations when the receivers with NLOS channels are known (K) or unknown (U). The gap between the performance when the NLOS receivers are known (dash-dotted line) and are unknown (solid line) represents the performance penalty for outlier identification.}\label{fig:NLOSidentloss}%
  \end{figure}

  The algorithm performs very close to the optimal ML estimator for the entire range of AoA spreads. However, it is interesting to note in Figure \ref{fig:NLOSidentloss} that the ML estimate does perform significantly worse compared to the ML estimate using only the good LOS AoA estimates. This additional loss is the cost of identifying the NLOS receivers and as the angular spread increases, it becomes progressively more difficult to differentiate between the LOS and NLOS AoA estimates.

\begin{figure}[ht]
  \centering
  \includegraphics[width=3.5in]{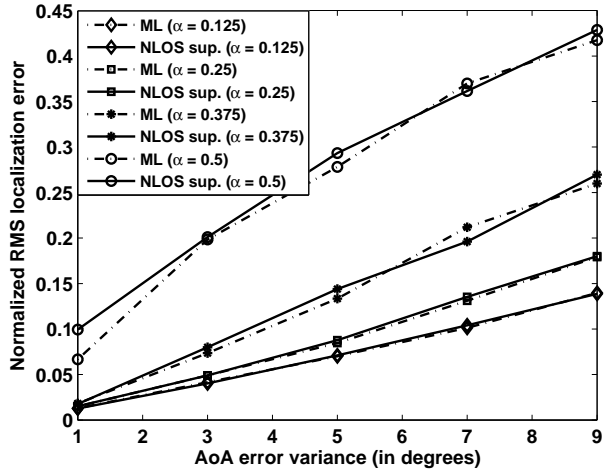}%
  \caption{Localization performance with NLOS suppression for an $N=8$ system in a realistic scenario with threshold chosen with the maximum expected outlier fraction, $\alpha_\trm{max} = 0.5$ for different fractions, $\alpha$, of receivers with NLOS channels. The location error variance (solid line) is compared against the ML error variance (dash-dotted line).}%
  \label{fig:NLOSlocerr_real}
\end{figure}

  In practice, since the fraction of outlying receivers, $\alpha$, is unknown, the algorithm is operated with a threshold chosen using an upper bound on this fraction, which in our simulations is $\alpha_\trm{max} = 0.5$. In Figure \ref{fig:NLOSlocerr_real}, for different numbers ($\{0,1,2,3,4\}$) of NLOS receivers, the outlier suppression algorithm achieves close to ML performance for smaller actual fractions of outliers even for this conservative choice of threshold. Thus, we can conclude that this approach is quite insensitive to the exact choice of threshold, $\Theta_\trm{max}$, which adds to the robustness of this approach. The simulation results also indicate that the algorithm in Section \ref{sec:aggr_outlierdet} is approximately ML (AML).

\begin{figure}[ht]
\centering
\includegraphics[width=3.5in]{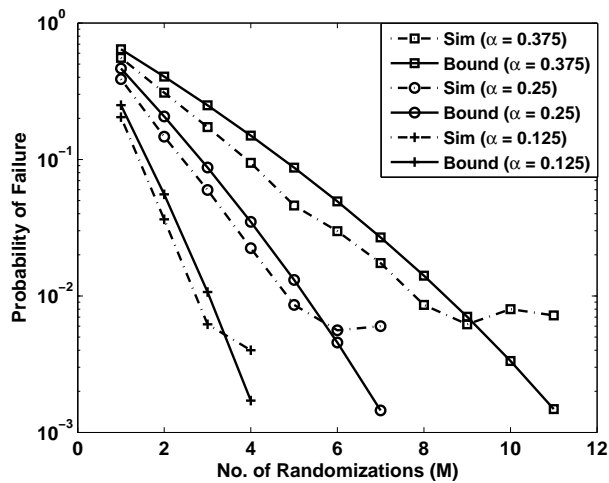}%
\caption{The comparison between the observed probability of failure and the expected probability due to bootstrap failures alone for a system with $N=8$ receivers with different fractions of outliers $\alpha$. The algorithm is defined to have ``failed'' if the SLE lies farther than three times the standard deviation of the ML estimator from the true location.}\label{fig:Probfail}%
\end{figure}

  In Figure \ref{fig:Probfail}, the observed probability of failure and the expected probability of failure from \eqref{eqn:boot_failure} of the outlier suppression algorithm are plotted for $N=8$ receivers. A failure is declared if the SLE is outside a circle of radius three times the standard deviation of the ML algorithm. When the SLE errors are normally distributed $N(0,\nu^2)$, as is expected from our analysis of the sequential algorithm in Section \ref{sec:props}, the rms localization error is $\textnormal{Rayleigh}(\nu)$. Thus, the probability of AoA measurement ``noise" alone causing the estimate to lie outside the $3\nu$ circle is $\exp{(-(3\nu)^2/\nu^2)} \approx 10^{-4}$. Hence, failures due to bad bootstraps are significant only when the observed failure probability is of the order of $10^{-3}$ and above. We observe from the figure that the expected failure probability is greater than the observed probability over almost the entire range of interest, but plateaus around $3\times 10^{-3}$ for all three fractions of outliers. This, we believe, is due to the fact that localization errors are not strictly normal in the presence of outliers leading to a slightly higher failure rate due to AoA measurement ``noise". We can safely conclude, therefore, that bootstrap failures dominate the failure probability over the entire range of interest.

  \subsubsection{Effect of alternate narrowband multipath models} 
\begin{figure}[ht]
\centering
\includegraphics[width=3.5in]{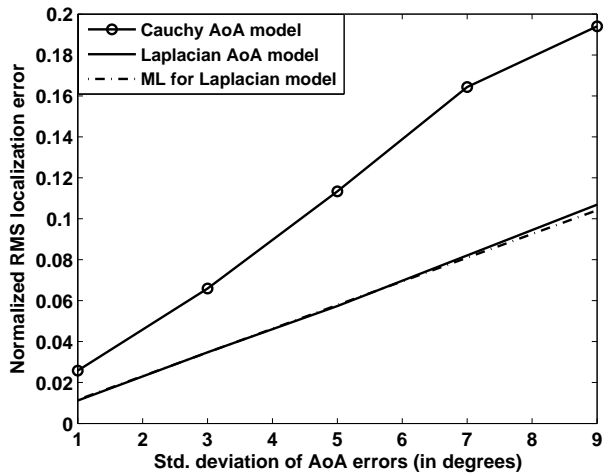}%
\caption{Localization performance of NLOS suppression when AoA estimates are obtained from Laplacian and Cauchy models with threshold chosen with the maximum expected outlier fraction, $\alpha_\trm{max} = 0.5$, as before.}\label{fig:NLOSothermods}%
\end{figure}

Although our algorithm is AML under the LOS and narrowband multipath models developed in Section \ref{sec:models}, it is of interest to examine the sensitivity of the NLOS suppression algorithm to these models. To this end, the algorithm was simulated, shown in Figure \ref{fig:NLOSothermods}, with two heavy-tailed AoA error models, namely the Laplacian (in \eqref{eqn:Lapmodel}) and Cauchy (the standard deviation corresponds to the shape parameter here), for the AoA estimation errors with the nominal parameters values in Figure \ref{fig:NLOSlocerr_real}; we did not compute the ML solution for the Cauchy model as it is mathematically intractable. Under the Laplacian model, which has smaller tails, the algorithm attained the optimal ML performance for that model. But, with a Cauchy model, the heavier tail generates many more outlying AoA estimates leading to larger estimation errors, and the observed performance is equivalent to that of our nominal model in \eqref{eqn:narrowbndmod} for $\alpha = 0.375$.

\subsubsection{Ray-tracing Illustrations}\label{sec:raytrace}
  We now compare the performance of the outlier suppression algorithm in the narrowband and wideband setting with the following two examples. In both examples, we use a virtual point source (shown as a solid circle) model to trace multipath generated by the reflectors; the LOS path from the virtual source to a receiver corresponds to the nominal direction of arrival of the reflected signal from the true source. In all examples, AoA estimation errors of standard deviation $0.5^\circ$ are added to both the nominal direction of arrival of the LOS path and the multipath from the virtual source.

\begin{figure}[ht]
  \includegraphics[width=3.5in]{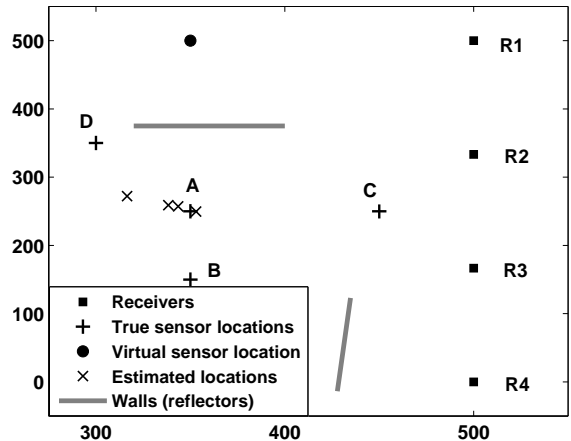}
  \centering
  \caption{The multipath suppression capability of the algorithm in a narrowband setting is illustrated with an example. With the source at location A, receiver R4 experiences LOS blockage and receives only reflected multipath from the source. The output of multiple runs of the outlier suppression algorithm with $M=7$ is shown for location A only. Location B uses only 3 receivers as R4 does not receive any signal from the source, location C uses all the receivers with LOS and location D has similar ray traces to location A. The rms error at these four different locations under this configuration are $18.42$ at A, $2.7$ at B, $2.57$ at C and $14.22$ at D (individual estimated locations not shown for B, C and D).}\label{fig:eg_2}
\end{figure}

 \textit{Narrowband multipath setting:}  In Figure \ref{fig:eg_2}, four receivers (squares) attempt to locate source A (`+' sign) in the presence of two reflectors (solid gray lines). One wall blocks the LOS path to receiver R4, causing R4 to only receive multipath reflected by the second wall. The receivers R1 and R2 have LOS AoA estimates, while receiver R3 receives both the LOS path and multipath from the virtual source (i.e., reflecting wall). The narrowband receivers generate one AoA estimate each, corresponding to the superposition of all the arriving paths. Thus, receivers R1 and R2 have very reliable LOS AoA estimates, R3 estimates an AoA with a large spatial spread and receiver R4 `sees' an outlying AoA measurement. The SLEs from multiple runs (crosses) of the outlier suppression algorithm is shown in Figure \ref{fig:eg_2}. The availability of reliable LOS estimates from R1 and R2 produces good estimates of the source location by eliminating the outlying estimate from R4 and also prevents the algorithm from mistaking the virtual source to be a real source. Localization performance at three other locations, B,C and D, is also shown in Figure \ref{fig:eg_2}. Good performance can, therefore, be expected in the narrowband setting even in the presence of LOS blockage if there are sufficiently many receivers with LOS paths to the source. However, there are situations, like when all the receivers experience NLOS propagation, where the narrowband system performs poorly. We elaborate on this issue in the following example.

\begin{figure}
\centering
\includegraphics[width=3.5in]{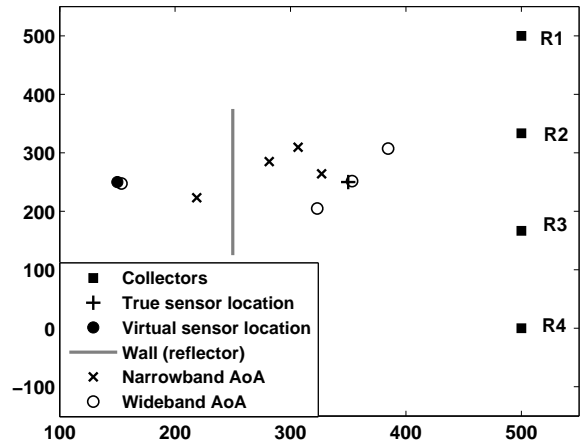}%
\caption{The working of the localization algorithm in the presence of a virtual source due to a perfect reflector (wall) with wideband and narrowband source transmissions. The output of multiple runs of the algorithm is shown.}\label{fig:eg_1}%
\end{figure}

\textit{Wideband multipath setting:} As described in Section \ref{sec:models}, when the source transmits a wideband signal, the receiver receivers can additionally resolve arriving paths in time, since reflected multipath components suffer delay with respect to the LOS path. This leads to multiple AoA estimates at the receiver corresponding to different directions and times of arrival. However, we cannot always reliably conclude that the earliest arriving path is LOS. Instead, we choose to apply our outlier suppression algorithm to all the estimated AoAs and allow the algorithm to eliminate NLOS AoA estimates as outliers. We illustrate this with an example in Figure \ref{fig:eg_1}. A source (`+' sign) is situated between four receivers (squares) and a wall (solid gray line). Under the wideband setting, in Figure \ref{fig:eg_1}, each receiver generates two AoA estimates, one due to the LOS path and the other due to the reflected NLOS path, while with the narrowband setting, the receivers estimate the AoA as a power-weighted superposition of the two directions of arrival. 

The output of multiple runs of the outlier suppression algorithm is plotted for the wideband and narrowband system. In the wideband system, the virtual source location is identified as a likely position in addition to the true source location and the NLOS algorithm is not capable of differentiating between true sources and virtual sources arising due to correlated multipath. But in practice, the knowledge of the environment can be used to eliminate infeasible estimates such as source locations behind the wall. On the contrary, in the narrowband scenario, the source is located in the region between the true and the virtual sources, as the multipath, in effect, increases the spatial spread in the LOS AoA estimates and degrades performance greatly. It is apparent that the capability to resolve multipath is essential to achieving satisfactory performance in NLOS environments and helps on two counts: First, resolving multiple incoming paths reduces the effective spatial spreading on each path. Second, multiple estimates at each receiver increase the total number of ``good" LOS measurements to estimate the source.

\section{Conclusions}\label{sec:concl}
The sequential algorithm nearly achieves the CRLB in an LOS environment, and forms the building block for our outlier suppression algorithm for NLOS environments.  Our algorithms have at most quadratic complexity in the number of measurements, are amenable to distributed implementation and can be easily modified to incorporate other sensing modalities such as RSS-based range measurements. For LOS environments, the scaling behavior of the localization error variance is linear in the AoA spatial spread (which determines the variance of the AoA estimates) and the size of the coverage area, and is inversely proportional to the number of receivers. Since we present results in terms of scale-invariant quantities, we can estimate the localization accuracy that can be obtained under any scenario as $(\textnormal{rms error from simulations}) \sqrt{\textnormal{Area covered}}/\pi $. Thus, for instance, in an outdoor environment of area 1 $\trm{km}^2$ with intermediate level of local scattering (we assume $4^\circ$ spatial spread from Table \ref{tab:aoaspreads}), the obtained localization accuracy using 8 receivers is about 15 m using simulation results in Figure \ref{fig:LOSrangeinfo}. On the other hand, with an angular spread of $9^\circ$ in the LOS path and $10\%$ outliers in an indoor WIMAX environment \cite{Akhdar2009}, an accuracy of 2 m can be obtained in a room of size 50x50 m using 8 receivers (see Figure \ref{fig:NLOSlocerr}). In NLOS multipath scenarios, although the outlier suppression algorithm is approximately ML, its performance is heavily dependent on the specific type of environment and the capability of the receivers to resolve the contributions of the LOS path and NLOS multipath in the received signal. In a narrowband system, where each receiver only resolves a superposition of the arriving paths, the algorithm can suppress outlying AoA estimates, if there are sufficient number of receivers with reliable LOS estimates. However, the capacity of a wideband system to estimate AoA from the LOS and multipath individually is vital in settings where all the receivers experience NLOS propagation. There is a localization performance penalty for having to ``find" the outliers, which becomes progressively worse as the fraction of outliers in the measurements increases. 

Broadly speaking, the key issue for further investigation is to relate physical models of the transceivers and the propagation environment to the AoA estimation models that form the basis for our algorithms and analysis.  This includes understanding the effect of the antenna array geometry, the placement of the receivers, and the propagation environment.  In particular, for canonical environments of interest (e.g. flat outdoor terrain, urban environments, indoor office environments, warehouses), it is of interest to develop models, given a deployment of receivers, for SNR variations at the receivers (due to variations in their distances from the source and multipath fading) and the fraction of receivers with blocked paths.  This in turn would provide a framework for optimizing the deployment of receivers.

\section*{Acknowledgments}
This work was supported by the National Science Foundation under grants CCF-0431205 and CNS-0520335, by the Office of Naval Research under grant N00014-06-0066, and by the Institute for Collaborative Biotechnologies under grant DAAD19-03-D-0004 from the US Army Research Office.

\appendix 
In this appendix, we derive the bounds on the probability of the bootstrap failure in \eqref{eqn:boot_failure}. As defined earlier, $P = \binom{N}{2}$ is the total number of bootstrap pairs and $K = P - \binom{\lfloor (1-\alpha) N\rfloor}{2}$ is the number of pairs with at least one NLOS receiver, where $\lfloor (1-\alpha) N\rfloor$ is the number of receivers with LOS channels and $\alpha$ is the fraction of receivers with NLOS channels.

\begin{eqnarray}
  P(\trm{bootstrap failure}) &=& \frac{\binom{M}{K}}{\binom{P}{M}} ~~\trm{if}~~ M\leq K \nonumber\\
                             &=& \left(1+\frac{K-P}{P}\right)\left(1+\frac{K-P}{P-1}\right)\ldots \left(1+\frac{K-P}{P-M+1}\right).\label{eqn:boot_fail_adix}
\end{eqnarray}
In order to obtain an upper bound on the bootstrap failure probability, we replace each ratio of the form $\frac{K-P}{P-k}$ by a larger fraction $\frac{K-P}{P}$ (note that $K<P$ by definition):
\begin{equation}
 P(\trm{bootstrap failure})\leq \left(\frac{K}{P}\right)^M.\label{eqn:bootupbnd}
\end{equation}
Similarly, replacing the ratios $\frac{K-P}{P-k}$ by a smaller fraction $\frac{K-P}{P-M+1}$ in \eqref{eqn:boot_fail_adix}, we get a lower bound,
\begin{equation}
 P(\trm{bootstrap failure})\geq \left(\frac{K-M+1}{P-M+1}\right)^M.\label{eqn:lowbnd}
\end{equation}
The lower and upper bounds clearly converge if $M\ll K<P$, which occurs when $N$ is large. Moreover, these bounds hold only for $M\leq K$, as the probability of bootstrap failure is zero if $M>K$.

\bibliographystyle{IEEEtran}
\bibliography{IEEEabrv,sensordriven}

\begin{thebibliography}{10}
\providecommand{\url}[1]{#1}
\csname url@samestyle\endcsname
\providecommand{\newblock}{\relax}
\providecommand{\bibinfo}[2]{#2}
\providecommand{\BIBentrySTDinterwordspacing}{\spaceskip=0pt\relax}
\providecommand{\BIBentryALTinterwordstretchfactor}{4}
\providecommand{\BIBentryALTinterwordspacing}{\spaceskip=\fontdimen2\font plus
\BIBentryALTinterwordstretchfactor\fontdimen3\font minus
  \fontdimen4\font\relax}
\providecommand{\BIBforeignlanguage}[2]{{%
\expandafter\ifx\csname l@#1\endcsname\relax
\typeout{** WARNING: IEEEtran.bst: No hyphenation pattern has been}%
\typeout{** loaded for the language `#1'. Using the pattern for}%
\typeout{** the default language instead.}%
\else
\language=\csname l@#1\endcsname
\fi
#2}}
\providecommand{\BIBdecl}{\relax}
\BIBdecl

\bibitem{Ananthasubramaniam2007}
B.~Ananthasubramaniam and U.~Madhow, ``Collector receiver design for data
  collection and localization in sensor-driven networks,'' in \emph{Proc.
  Conference on Information Sciences and Systems, 2007.}, 2007, pp. 591--596.

\bibitem{aeroscout}
\BIBentryALTinterwordspacing
``Aeroscout inc.'' [Online]. Available: \url{http://www.aeroscout.com/}
\BIBentrySTDinterwordspacing

\bibitem{savi}
\BIBentryALTinterwordspacing
``Savi technology inc.'' [Online]. Available:
  \url{http://www.savi.com/index.shtml}
\BIBentrySTDinterwordspacing

\bibitem{wherenet}
\BIBentryALTinterwordspacing
``Wherenet inc.'' [Online]. Available: \url{http://www.wherenet.com/}
\BIBentrySTDinterwordspacing

\bibitem{ekahau}
\BIBentryALTinterwordspacing
``Ekahau inc.'' [Online]. Available: \url{http://www.ekahau.com/}
\BIBentrySTDinterwordspacing

\bibitem{melt08}
B.~Ananthasubramaniam and U.~Madhow, ``Cooperative localization using angle of
  arrival measurements in non-line-of-sight environments,'' in \emph{MELT '08:
  Proc. ACM international workshop on Mobile entity localization and tracking
  in GPS-less environments}, San Francisco, CA, Sept. 2008, pp. 117--122.

\bibitem{Patwari2005}
N.~Patwari, J.~Ash, S.~Kyperountas, A.~Hero, R.~Moses, and N.~Correal,
  ``Locating the nodes: cooperative localization in wireless sensor networks,''
  \emph{{IEEE} Signal Process. Mag.}, vol.~22, no.~4, pp. 54--69, 2005.

\bibitem{Mao2007}
G.~Mao, B.~Fidan, and B.~D. Anderson, ``Wireless sensor network localization
  techniques,'' \emph{Computer Networks}, vol.~51, no.~10, pp. 2529--2553, Jul.
  2007.

\bibitem{robust}
R.~A. Maronna, D.~R. Martin, and V.~J. Yohai, \emph{Robust Statistics: Theory
  and Methods}.\hskip 1em plus 0.5em minus 0.4em\relax Wiley, 2006.

\bibitem{borras}
J.~Borras, P.~Hatrack, and N.~Mandayam, ``A decision theoretic framework for
  {NLOS} identification,'' in \emph{Proc. VTC'98}, vol.~2, May 1998, pp.
  1583--1587.

\bibitem{Cong2005}
L.~Cong and W.~Zhuang, ``Nonline-of-sight error mitigation in mobile
  location,'' \emph{{IEEE} Trans. Wireless Commun.}, vol.~4, no.~2, pp.
  560--573, 2005.

\bibitem{venkat}
S.~Venkatraman and J.~Caffery~Jr, ``Statistical approach to nonline-of-sight
  {BS} identification,'' in \emph{Proc. WPMC'02}, vol.~1, Honolulu, Hawaii, Oct
  2002, pp. 296--300.

\bibitem{bao}
B.~L. Le, K.~Ahmed, and H.~Tsuji, ``Mobile location estimator with {NLOS}
  mitigation using kalman filtering,'' in \emph{Proc. WCN'03}, vol.~3, Mar.
  2003, pp. 1969--1973.

\bibitem{seshan}
\BIBentryALTinterwordspacing
S.~Srirangarajan and A.~H. Tewfik, ``Sensor node localization via spatial
  domain quasi-maximum likelihood estimation,'' in \emph{Proc. EUSIPCO'06},
  2006. [Online]. Available:
  \url{http://www.arehna.di.uoa.gr/Eusipco2006/papers/1568979916.pdf}
\BIBentrySTDinterwordspacing

\bibitem{xiong}
L.~Xiong, ``A selective model to suppress {NLOS} signals in angle-of-arrival
  ({AOA}) location estimation,'' in \emph{Proc. PIMRC'98}, vol.~1, Sept. 1998,
  pp. 461--465.

\bibitem{Yu2009}
K.~Yu and Y.~Guo, ``Statistical {NLOS} identification based on {AOA}, {TOA},
  and signal strength,'' \emph{{IEEE} Trans. Veh. Technol.}, vol.~58, no.~1,
  pp. 274--286, 2009.

\bibitem{Guevenc2008}
I.~G\"{u}ven\c{c}, C.-C. Chong, F.~Watanabe, and H.~Inamura, ``{NLOS}
  identification and weighted least-squares localization for {UWB} systems
  using multipath channel statistics,'' \emph{EURASIP Journal on Advances in
  Signal Processing}, vol. 2008, p.~14, 2008, article ID 271984, doi:
  10.1155/2008/271984.

\bibitem{Al-Jazzar2009}
S.~Al-Jazzar, M.~Ghogho, and D.~McLernon, ``A joint {TOA}/{AOA} constrained
  minimization method for locating wireless devices in non-line-of-sight
  environment,'' \emph{{IEEE} Trans. Veh. Technol.}, vol.~58, no.~1, pp.
  468--472, 2009.

\bibitem{Guvenc2009}
I.~{G\"{u}ven\c{c}} and C.-C. Chong, ``A survey on {TOA} based wireless
  localization and {NLOS} mitigation techniques,'' \emph{{IEEE} Commun. Surveys
  Tuts.}, 2009, to appear.

\bibitem{Tang2008}
H.~Tang, Y.~Park, and T.~Qiu, ``A {TOA}-{AOA}-based {NLOS} error mitigation
  method for location estimation,'' \emph{EURASIP Journal on Advances in Signal
  Processing}, vol. 2008, p.~14, 2008, article ID 682528,
  doi:10.1155/2008/682528.

\bibitem{chen}
P.~C. Chen, ``A non-line-of-sight error mitigation algorithm in location
  estimation,'' in \emph{Proc. WCNC'99}, Sept. 1999, pp. 316--320.

\bibitem{Venkatesh2007}
S.~Venkatesh and R.~Buehrer, ``{NLOS} mitigation using linear programming in
  ultrawideband location-aware networks,'' \emph{{IEEE} Trans. Veh. Technol.},
  vol.~56, no.~5, pp. 3182--3198, 2007.

\bibitem{casas}
R.~Casas, A.~Marco, J.~J. Guerrero, and J.~Falc\'{o}, ``Robust estimator for
  non-line-of-sight error mitigation in indoor localization,'' \emph{EURASIP
  Journal on Applied Signal Processing}, vol. 2006, p.~8, 2006,
  doi:10.1155/ASP/2006/43429.

\bibitem{godara}
L.~C. Godara, ``Application of antenna arrays to mobile communications. {II}.
  beam-forming and direction-of-arrival considerations,'' \emph{Proc. {IEEE}},
  vol.~85, no.~8, pp. 1195--1245, Aug 1997.

\bibitem{raich}
R.~Raich, J.~Goldberg, and H.~Messer, ``Bearing estimation for a distributed
  source: Modeling, inherent accuracy limitations, and algorithms,''
  \emph{{IEEE} Trans. Signal Process.}, vol.~48, pp. 429--441, Feb. 2000.

\bibitem{ertel}
R.~B. Ertel, K.~Sowerby, T.~S. Rappaport, and J.~H. Reed, ``Overview of spatial
  channel models for antenna array communication systems,'' \emph{{IEEE}
  Personal Commun. Mag.}, pp. 10--22, 1998.

\bibitem{meng}
Y.~Meng, P.~Stoica, and K.~M.Wong, ``Estimation of the directions of arrival of
  spatially dispersed signals in array processing,'' \emph{IEE Proceedings
  Radar, Sonar and Navigation}, vol. 143, no.~1, pp. 1--9, Feb. 1996.

\bibitem{trump}
T.~Trump and B.~Ottersten, ``Estimation of nominal direction of arrival and
  angular spread using an array of sensors,'' \emph{Signal Processing},
  vol.~50, no. 1-2, pp. 57--69, 1996.

\bibitem{valaee}
S.~Valaee, B.~Champagne, and P.~Kabal, ``Parametric localization of distributed
  sources,'' \emph{{IEEE} Trans. Signal Process.}, vol.~43, no.~9, pp.
  2144--2153, Sept. 1995.

\bibitem{spencer}
Q.~H. Spencer, B.~D. Jeffs, M.~A. Jensen, and A.~L. Swindlehurst, ``Modeling
  the statistical time and angle of arrival characteristics of an indoor
  multipath channel,'' \emph{{IEEE} J. Sel. Areas Commun.}, vol.~18, no.~3, pp.
  347--360, Mar. 2000.

\bibitem{rao_hari}
B.~Rao and K.~Hari, ``Performance analysis of root-music,'' \emph{{IEEE} Trans.
  Signal Process.}, vol.~37, no.~12, pp. 1939--1949, Dec 1989.

\bibitem{GPS}
B.~W. Parkinson and J.~J. Spilker, \emph{Global Positioning System: Theory and
  Practice}.\hskip 1em plus 0.5em minus 0.4em\relax American Institute of
  Aeronautics, 1996.

\bibitem{Coleman1996}
T.~F. Coleman and Y.~Li, ``An interior trust region approach for nonlinear
  minimization subject to bounds,'' \emph{SIAM Journal on Optimization},
  vol.~6, no.~2, pp. 418--445, 1996.

\bibitem{poor_book}
H.~V. Poor, \emph{An Introduction to Signal Detection and Estimation},
  2nd~ed.\hskip 1em plus 0.5em minus 0.4em\relax Springer, 2005.

\bibitem{Klein1996}
A.~Klein, W.~Mohr, R.~Thomas, P.~Weber, and B.~Wirth, ``Direction-of-arrival of
  partial waves in wideband mobile radio channels for intelligent antenna
  concepts,'' in \emph{Proc. VTC}, vol.~2, 1996, pp. 849--853.

\bibitem{Thomas1992}
H.~J. Thomas, T.~Ohgane, and M.~Mizuno, ``A novel dual antenna measurement of
  the angular distribution of received waves in the mobile radio environment as
  a function of position and delay time,'' in \emph{Proc. VTC}, vol.~1, 1992,
  pp. 546--549.

\bibitem{Spencer1997}
Q.~Spencer, M.~Rice, B.~Jeffs, and M.~Jensen, ``Indoor wideband time/angle of
  arrival multipath propagation results,'' in \emph{Proc. VTC}, vol.~3, 1997,
  pp. 1410--1414.

\bibitem{Akhdar2009}
O.~Akhdar, D.~Carsenat, C.~Decroze, M.~Mouhamadou, and T.~Monediere,
  ``Direction of arrival measurements for outdoor-to- indoor channel
  characterization,'' in \emph{Proc. EuCAP 2009}, Mar. 2009, pp. 2280--2282.

\bibitem{Chen2001}
M.~Chen and H.~Asplund, ``Measurements and models for direction of arrival of
  radio waves in los in urban microcells,'' in \emph{Proc. PIMRC}, vol.~1,
  2001, pp. B--100-- B--104.

\bibitem{Patwari2003}
N.~Patwari, I.~Hero, A.O., M.~Perkins, N.~Correal, and R.~O'Dea, ``Relative
  location estimation in wireless sensor networks,'' \emph{{IEEE} Trans. Signal
  Process.}, vol.~51, no.~8, pp. 2137--2148, Aug. 2003.

\end{thebibliography}
%
%\noindent {\bf\large Supplementary Material}\\
% The performance of the sequential algorithm in Section \ref{sec:localg} for different numbers of receivers is shown in Figure \ref{fig:LOSlocerr}. The algorithm achieves the CRLB for small angular estimation errors even for as few as 6 receivers.
%\begin{figure}[ht]
% \centering
%    \includegraphics[width=3.5in]{LOSlocerr}%
%    \caption{Localization performance of the sequential estimation algorithm (in dashed lines) in the LOS scenario for $N = 6,8,12$ receivers. The performance is compared against the CRLB (in solid lines) and the ML estimator (in dotted lines).}\label{fig:LOSlocerr}%
%\end{figure}
%\begin{figure}[ht]
%\centering
%  \includegraphics[width=3.5in]{Nscale}%
%  \caption{Log-Log plot of the mean square localization error against number of receivers $N$ using the sequential algorithm with AoA standard deviation of $1^{o}$. The Cramer-Rao bound is also plotted alongside.}\label{fig:Nscale}%
%\end{figure}
%
%The localization accuracy can also be improved by the addition of receivers to the system. In Figure \ref{fig:Nscale}, we use a log-log plot to study the dependence of the localization error on the number of receivers $N$. A linear fit of the simulated data shows that the resolution scales linearly with $N-1$, which agrees with the dependence deduced from the Cramer Rao bound in \eqref{eqn:centerbound}. The resolution scales as $N-1$ and not $N$ is due the fact that at least two receivers are needed to locate a source and we receive the benefit of error averaging only over the remaining $N-1$ AoA estimates.
\end{document}